\documentclass[12pt]{JHEP3}
\usepackage{amsmath,amssymb,amsfonts,amsthm,mathrsfs}
\usepackage{bm}
\usepackage{graphicx}

\newcommand{\be}{\begin{equation}}
\newcommand{\ee}{\end{equation}}
\newcommand{\ba}{\begin{eqnarray}}
\newcommand{\ea}{\end{eqnarray}}
\def\bs{\begin{subequations}}
\def\es{\end{subequations}}
\def\a{\alpha}

\def\g{\gamma}
\def\k{\kappa}
\def\e{\epsilon}
\newcommand{\V}{{\text{\tiny $V$}}}
\newcommand{\f}{{\text{\tiny f}}}

\def\s{\sigma}

\def\vp{\varphi}
\def\dpl{\delta_{\rm Pl}}

\def\cH{\mathcal{H}}

\def\cV{{\cal V}}

\newcommand{\Eq}[1]{(\ref{#1})}

\def\rme{\text{e}}
\def\rmd{\text{d}}
\def\rmi{\text{i}}

\title{Observational test of inflation in loop quantum cosmology}

\author{Martin Bojowald$^{1}$, Gianluca Calcagni$^{2}$, 
and Shinji Tsujikawa$^{3}$\\
$^1$Institute for Gravitation and the Cosmos, The Pennsylvania State University\\
104 Davey Lab, University Park, PA 16802, U.S.A.\\
$^2$Max Planck Institute for Gravitational Physics (Albert Einstein Institute)\\ 
Am M\"uhlenberg 1, D-14476 Golm, Germany\\
$^3$Department of Physics, Faculty of Science, Tokyo University of Science\\
1-3, Kagurazaka, Shinjuku-ku, Tokyo 162-8601, Japan\\
E-mail: \email{bojowald@gravity.psu.edu}, \email{calcagni@aei.mpg.de}, \email{shinji@rs.kagu.tus.ac.jp}}

\date{July 8, 2011}

\abstract{
We study in detail the power spectra of scalar and tensor perturbations generated during inflation in loop quantum cosmology (LQC). After clarifying in a novel quantitative way how inverse-volume corrections arise in inhomogeneous settings, we show that they can generate large running spectral indices, which generally lead to an enhancement of power at large scales. We provide explicit formul\ae\ for the scalar/tensor power spectra under the slow-roll approximation, by taking into account corrections of order higher than the runnings. Via a standard analysis, we place observational bounds on the inverse-volume quantum correction $\delta \propto a^{-\sigma}$ ($\sigma>0$, $a$ is the scale factor) and the slow-roll parameter $\epsilon_\V$ for power-law potentials as well as exponential potentials by using the data of WMAP 7yr combined with other observations. We derive the constraints on $\delta$ for two pivot wavenumbers $k_0$ for several values of $\delta$. The quadratic potential can be compatible with the data even in the presence of the LQC corrections, but the quartic potential is in tension with observations. We also find that the upper bounds on $\delta (k_0)$ for given $\sigma$ and $k_0$ are insensitive to the choice of the inflaton potentials.
}


\preprint{\small Journal of Cosmology and Astroparticle Physics 11 (2011) 046,\qquad\qquad arXiv:1107.1540}

\begin{document}


%
\section{Introduction}

Both the construction of quantum gravity and the question of its observational tests
are beset by a host of problems. On the one hand,
quantum gravity, in whatever approach, must face many mathematical
obstacles before it can be completed to a consistent theory. On the
other hand, assuming that a consistent theory of quantum gravity does
exist, dimensional arguments suggest that its observational
implications are of small importance. In the realm of cosmology, for instance, they are estimated to be of
the tiny size of the Planck length divided by the Hubble distance. 

Between the two extremes of conceptual
inconsistency and observational irrelevance lies a window of
opportunity in which quantum gravity is likely to fall. It is true
that we do not yet know how to make quantum gravity fully consistent,
and it is true that its effects for early-universe cosmology should be
expected to be small. But in trying to make some quantum-gravity
modified cosmological equations consistent, it has been found that
there can be stronger effects than dimensional arguments
suggest. Consistency requirements, especially of loop quantum gravity,
lead to modified spacetime structures that depart from the usual
continuum, implying unexpected effects.

To partially bridge the gap between fundamental developments and
loop quantum gravity phenomenology and observations, one considers
effective dynamics coming from constraint functions evaluated on
a particular background and on a large class of semiclassical
states. In generally-covariant systems, the dynamics is fully
constrained, and the constraint functionals on phase space generate
gauge transformations obeying an algebra that reveals the
structure of spacetime deformations. The algebra of these gauge
generators thus shows what underlying notion of spacetime covariance
is realized, or whether covariance might be broken by quantum effects,
making the theory inconsistent. If a consistent version with an
unbroken (but perhaps deformed) gauge algebra exists, it can be
evaluated for potential observational implications.

These effective constraints and their algebra, at the present
stage of developments, can be evaluated for perturbative
inhomogeneities around a cosmological background. While loop quantum
gravity is background independent in the sense that no spacetime
metric is assumed before the theory is quantized, a background and the
associated perturbation theory can be introduced at the effective
level. In the case of interest here, the background is cosmological, a
flat Friedmann--Robertson--Walker (FRW) spacetime with a rolling
scalar field and perturbed by linear inhomogeneous fluctuations of the
metric-matter degrees of freedom. The idea is to implement as many
quantum corrections as possible, and study how the inflationary
dynamics is modified. In this context, it is important that no gauge
fixing be used before quantization, as such a step would
invariably eliminate important consistency conditions by fiat, not by
solving them. The result would be a framework whose ``predictions''
depend on how the gauge was fixed in the first place.

This program, challenging as it is, has been carried out only
partially so far, and in gradual steps. First, the full set of
constraints was derived for vector \cite{BH1}, tensor \cite{BH2}, and
scalar modes \cite{BHKS,BHKS2} in the presence of small inverse-volume
corrections. The gravitational wave spectra have been studied in
\cite{CMNS2,CH}, where the effect of inverse-volume corrections and
their observability were discussed. The scalar inflationary spectra
and the full set of linear-order cosmological observables was then
derived in \cite{InflObs}, thus making it possible to place
observational bounds on the quantum corrections themselves
\cite{BCT} (see also refs.~\cite{inflationworks} for early related works). 
The reason why these studies concentrate on inverse-volume
corrections is mainly technical; in fact, the closure of the
constraint algebra has been verified only in this case (and only in
the limit of small corrections).  A class of consistent constraints
with a closed algebra is known also for vector and tensor modes in the
presence of holonomy corrections \cite{BH1,BH2,MCBG}, and since recently also for the scalar sector, where anomaly cancellation is more difficult to work out \cite{Bar}. Inspections of cosmological holonomy effects have so far been limited to the tensor sector \cite{CMNS2,Mie1,GB1,GB2,GB3}.

Exactly isotropic minisuperspace models, where the situation is
reversed and holonomy corrections are easier to implement than
inverse-volume corrections, provide another reason why it is
interesting to focus attention on inverse-volume
corrections. Effective equations available for certain matter contents
with a dominating kinetic energy \cite{BouncePert,HighDens} suggest
that holonomy corrections are significant only in regimes of
near-Planckian densities \cite{APSII} but not during the timespan
relevant for early-universe cosmology, including inflation.
Inverse-volume corrections, on the other hand, do not directly react
to the density but rather to the discreteness scale of quantum
gravity, which is not determined immediately by the usual cosmological
parameters. The question of whether they are small or can play a
significant role must be answered by a self-consistent
treatment.

Such a treatment shows that inverse-volume corrections present an
example of quantum-gravity effects that can be larger than what
dimensional arguments suggest \cite{BCT}.  Here we present the full
details of the analysis briefly reported in \cite{BCT} for a quadratic
inflationary potential, enriching it with new constraints on other
potentials. From a cosmological perspective, we shall provide the
complete set of slow-roll equations as functions of the potential,
extend the likelihood analysis to quartic and exponential potentials, and discuss how
the experimental pivot scale and cosmic variance affect the results.

Before examining the details and experimental bounds of the model,
from a quantum-gravity perspective we will clarify some conceptual
issues which must be taken into account for a consistent treatment of
inverse-volume corrections. In particular, we justify for the first
time why inverse-volume corrections depend only on triad variables,
and not also on connections. Until now, this was regarded as a
technical assumption devoid of physical motivations. Here we show it
as a consequence of general but precise semiclassical
arguments. Further, we spell the reason why inverse-volume corrections
are not suppressed at the inflationary density scale.

The plan of the paper is as follows. We begin in section \ref{invo} by
discussing inverse-volume corrections in LQC and their justification
in inhomogeneous models, providing the first implementation for a
class of semiclassical states sufficiently large to be used in
effective equations. The lattice refinement picture has been introduced and developed in a number of papers, but a major twofold open issue remains. On one hand, there is the need to justify why inverse-volume corrections depend only on triad variables and not on holonomies. On the other hand, it is not clear how the fundamental discreteness scale, previously introduced \emph{in media res}, arises in the lattice refinement framework. Section \ref{invo} does both systematically for the first time. The relation between (the size of) inverse-volume and holonomy quantum corrections is clarified in section \ref{23}, while section \ref{24} is a recapitulation of past criticism and a discussion of how it is addressed by the present arguments. After that, we turn to an application for
observational cosmology. In section \ref{hsr} we review the formul\ae\ of the
cosmological observables in the Hubble slow-roll tower
\cite{InflObs}. In section \ref{vsr} we reexpress these quantities in the
slow-roll parameters as functions of the inflationary potential. In
section \ref{ksr} the observables are recast as functions of the momentum
pivot scale for a ready use in numerical programs. The effect of
cosmic variance on the scalar power spectrum is also discussed
therein. In section \ref{likelihoodsec} we shall carry out the likelihood analysis 
to constrain the inverse-volume corrections in the presence of
several different inflaton potentials by using the observational 
data of Cosmic Microwave Background (CMB) combined with 
other datasets.

\section{Cosmology with a discrete scale}\label{invo}

One of the main features of loop quantum gravity (LQG), shared with
other approaches to quantum gravity, is the appearance of discrete
spatial structures replacing the classical continuum of general
relativity. It is often expected that the scale of the discreteness is
determined by the Planck length $\ell_{\rm Pl}=\sqrt{G\hbar}$, but if
discreteness is fundamental, its scale must be set by a dynamical
parameter of some underlying state, just as the lattice spacing of a
crystal is determined by the interaction of atoms. 
In this section, we develop the cosmological picture of dynamics of discrete
space, highlighting the form of quantum corrections to be
expected. Readers more interested in potentially observable
consequences may skip this technical part, but those acquainted with LQC will 
find a fresh discussion and justification of inverse-volume corrections
and the lattice refinement picture.

In loop quantum gravity, such states are represented by spin networks,
graphs in an embedding space whose edges $e$ are labeled by spin
quantum numbers $j_e$. The quantum number determines the area of an
elementary plaquette intersecting only one edge $e$, given by ${\cal
A}=\gamma \ell_{\rm Pl}^2 \sqrt{j_e(j_e+1)}$. As the plaquette is
enlarged, its geometrical size changes only when it begins to
intersect another edge, increasing in quantum jumps. In the area
formula, $\gamma$ is the Barbero--Immirzi parameter, whose size
(slightly less than one) can be inferred from computations of
black-hole entropy \cite{ABCK,ABK}. As expected, the scale is set by
the Planck length for dimensional reasons, but the actual size is
given by the spin quantum number. Its values in a specific physical
situation have to be derived from the LQG dynamical equations, a task
which remains extremely difficult. However, given the form in which
$j_e$ appears in the dynamical equations, its implications for physics
can be traced and parametrized in sufficiently general form so as to
analyze effects phenomenologically.

\subsection{Scales}

In order to model this situation in general terms, we begin with a
nearly isotropic spacetime and a chunk of space of some comoving size
${\cal V}_0$, as measured by the extensions in some set of
coordinates. The geometrical size is then $\cV={\cal V}_0 a^3$, where
$a$ is the scale factor. We complement this classical picture with a
discrete quantum picture, in which the same chunk of space is made up
from nearly-isotropic discrete building blocks, all of the same size
$v$. If there are ${\cal N}$ discrete blocks in a region of size
${\cal V}$, we have the relationship $v={\cal V}_0 a^3/{\cal N}$. The
elementary volume $v$, or the linear scale $L=v^{1/3}$, will be our
main parameter, related to the quantum state via its labels $j_e$. The
elementary quantum-gravity scale $L$ need not be exactly the Planck
length, depending on what $j_e$ are realized. Instead of using the
$j_e$, which are subject to complicated dynamics, it turns out to be
more useful to refer to $L$ in phenomenological
parametrizations. Similarly, we define the quantum-gravity density
scale
\begin{equation}
\rho_\textsc{qg}= \frac{3}{8\pi G L^2}\,,
\end{equation}
which equals $3/8\pi$ times the Planck density for $L=\ell_{\rm Pl}$.

In loop quantum gravity, the discreteness is mathematically seen as a
rather direct consequence of the fact that the fundamental operators
are holonomies along curves $e$, computed for a certain form of
gravitational connection, the Ashtekar--Barbero connection
$A_a^i$,\footnote{Indices $a,b,\dots=1,2,3$ run over space directions,
while $i,j,\dots=1,2,3$ are internal indices in the su(2) algebra.}
while the connection itself is not a well-defined operator. For a
nearly isotropic spacetime, there is only one nontrivial connection
component, given by $c=\gamma\dot{a}$ in terms of the proper-time
derivative of the scale factor. The classical holonomies are
$h_e={\cal P}\exp(\int_e\rmd\lambda\,\dot{e}^aA_a^i\tau_i)$, with
$\tau_j=\rmi\sigma_j/2$ proportional to Pauli matrices and path
ordering indicated by ${\cal P}$. Every $h_e$ takes values in the
compact group SU(2), whose representations appear as the spin labels
of edges $j_e$, giving rise to discrete conjugate variables.

Another consequence of one being able to represent only holonomies,
not connection components, is that the usual polynomial terms in
connection-dependent Hamiltonians are replaced by the whole series
obtained by expanding the exponential expression for an holonomy. In
this way, higher-order corrections are implemented in the
dynamics. Corrections become significant when the argument of
holonomies, given by line integrals of $A_a^i$ along the spin-network
edges, is of order one. For a nearly isotropic connection
$A_a^i=c\delta_a^i$, the integral along straight lines reduces to
$\ell_0c$, where $\ell_0$ is the coordinate (i.e., comoving) length of
the edge. If the edge is elementary and of the discreteness size of
our underlying state, we have $\ell_0=L/a=v^{1/3}/a$, and the
condition for holonomy corrections becoming large is $v^{1/3}c/a\sim
1$. More intuitively, holonomy corrections become large when the
Hubble scale $H^{-1}= a/\dot{a}\sim \gamma L$ is of the size of the
discreteness scale, certainly an extreme regime in cosmology. Yet
another intuitive way of expressing this regime is via densities:
holonomy corrections are large when the matter density is of the order
of the quantum-gravity density. By the classical Friedmann equation,
this happens when
\begin{equation}
\rho=\frac{3}{8\pi G} H^2=\frac{3}{8\pi G}\frac{c^2}{a^2\gamma^2} \sim \frac{3}{8\pi
 G\gamma^2L^2}= \gamma^{-2}\rho_\textsc{qg}\,.
\end{equation}
We introduce the
parameter $\delta_{\rm hol}:= \rho/\rho_\textsc{qg}= 8\pi
G v^{2/3}\rho/3$ in order to quantify holonomy
corrections. These are small when $\delta_{\rm hol}\ll 1$.

The discreteness of loop quantum gravity manifests itself in different
ways, some of which require more details to be derived. In addition to holonomy 
corrections, the most
important one arises when one considers the inverse of the elementary lattice areas. Classically, the
areas correspond to the densitized triad $E^a_i$, which determines the
spatial metric $q_{ab}$ via $E^a_iE^b_i= q^{ab}\det q$ and is
canonically conjugate to the connection $A_a^i$. The inverse of
$E^a_i$ or its determinant appears in the Hamiltonian
constraint of gravity as well as in all the usual matter
Hamiltonians, especially in kinetic terms, and is thus crucial for the dynamics. 

Upon quantization, however, the densitized triad is represented in terms of
the spin labels that also determine the lattice areas, and those
labels can take the value zero. No densely defined inverses of the area
operators exist, and therefore there is no direct way to quantize
inverse triads or inverse volumes as they appear in
Hamiltonians. However, as with holonomies replacing connection
components, there is an indirect way of constructing well-defined
inverse-volume operators, 
which imply further quantum corrections.

\subsection{Derivation of inverse-volume corrections}

The quantization of different kinds of inverse volumes or the co-triad
$e_a^i$, obtained from the inverse of $E^a_i$, begins with Poisson
identities such as \cite{QSDI,QSDV}
\begin{equation} \label{cotriadPoisson}
 \left\{A_a^i,\int\rmd^3x\sqrt{|\det E|}\right\}= 2\pi\gamma G
 \epsilon^{ijk}\epsilon_{abc} \frac{E^b_jE^c_k}{\sqrt{|\det E|}}\,
{\rm sgn}(\det E)
=4\pi\gamma G e_a^i\,,
\end{equation}
stemming from the basic Poisson brackets $\{A^i_a(x),E^b_j(y)\}= 8\pi
G\gamma \delta^i_j\delta^b_a\delta(x,y)$.  On the right-hand side of
eq.~(\ref{cotriadPoisson}), there is an inverse of the determinant of
$E^a_i$, but on the left-hand side no such inverse
is required. Classically, the inverse arises from derivatives
contained in the Poisson bracket, but after quantization the Poisson
bracket is replaced by a commutator and no derivative or inverse
appears. In this way, one obtains well-defined operators for the inverse 
volume, implementing an automatic ultraviolet cutoff at small length scales. 

The volume $\int\rmd^3x\sqrt{|\det E|}$ of some region,
containing the point v where we want to evaluate the co-triad, is
quantized by well-defined volume operators, and the connection can be
represented in terms of holonomies. For holonomies with edges of comoving length $\ell_0$, 
we can write
\begin{equation} \label{commPoiss}
 {\rm tr}(\tau^ih_{{\rm v},e}[h_{{\rm v},e}^{-1},\hat{V}_{\rm v}])\sim \frac{1}{2}\rmi\hbar\ell_0
\widehat{\{A_a^i,V_{\rm v}\}}\dot{e}^a \,.
\end{equation}
Here, $\tau^j=\rmi\s^j/2$ are Pauli matrices, $h_{{\rm v},e}$ is a
holonomy starting at a lattice vertex v in some direction $e$, and
$V_{\rm v}$ is the volume of some region around v, with $\hat{V}_{\rm
v}$ its quantization. As long as v is included in the region
integrated over to obtain the volume, it does not matter how far the
region extends beyond v. One could even use the volume of the whole
space.

To quantize, loop quantum gravity provides the holonomy-flux
representation of the basic operators $\hat{h}_{{\rm v},e}$
(holonomies along edges $e$) and $\hat{F}_S=\int_S \rmd^2y\,E^a_in_a$,
fluxes of the densitized triad through surfaces $S$ with co-normal
$n_a$. These variables are SU(2)-valued, but one can devise a regular
lattice for a simple implementation of inhomogeneity, setting edges
with tangent vectors $\dot{e}^a_I=\delta_I^a$, $I=1,2,3$ in Cartesian
coordinates.  Then, holonomies are given by $h_{{\rm v},e_I}=
\exp(\ell_0\tau_Ic)= \cos(\ell_0c/2)+2\tau_I\sin(\ell_0c/2)\in {\rm
SU}(2)$, where $c$ is the connection evaluated somewhere on the
edge. All connection-dependent matrix elements can thus be expressed
in the complete set of functions $\eta:=\exp(\rmi\ell_0c/2)\in{\rm
U}(1)$, and the flux through an elementary lattice site in a nearly
isotropic geometry is simply $F=\ell_0^2 p$ with $|p|\sim a^2$, and
$p$ carrying a sign amounting to the orientation of space. Isotropy
thus allows a reduction from SU(2) to U(1), with certain technical
simplifications.

For a nearly isotropic configuration, we assign a copy of the
isotropic quantum theory to every (oriented) link $I$ of a regular
graph, making the theory inhomogeneous. By this step we certainly do
not reach the full theory of loop quantum gravity, which is based on
irregular graphs with SU(2)-theories on its links. But we will be able
to capture the main effects which have appeared in approximate
considerations of loop quantum gravity with simpler graphs and reduced
gauge groups.  The basic operators are then a copy of $\hat{\eta}_{{\rm v},I}$ and
$\hat{F}_{{\rm v},I}$ for each lattice link with \be\label{core}
[\hat{\eta}_{{\rm v},I},\hat{F}_{{\rm v}',J}]= -4\pi \gamma \ell_{\rm
Pl}^2\hat{\eta}_{{\rm v},I} \delta_{IJ}\delta_{{\rm v},{\rm v}'}\,, \ee if the
edge of the holonomy and the surface of the flux intersect.

When we insert holonomies for nearly isotropic connections in
eq.~(\ref{commPoiss}) and evaluate the trace, inverse-volume operators
resulting from commutators have the form
\begin{equation}
\hat B_{{\rm v},I} = \frac{1}{4\pi \gamma G\hbar} \left(\hat{\eta}_{{\rm v},I}^{\dagger}
 \hat V_{\rm v}\hat{\eta}_{{\rm v},I} - \hat{\eta}_{{\rm v},I}\hat V_{\rm v} \hat{\eta}_{{\rm v},I}^{\dagger}\right)\,.
\end{equation}
The volume at vertex v is obtained from components of the densitized
triad, quantized by a flux operator $\hat{F}_{{\rm v},I}$, with v an
endpoint of the link $I$. If $(I,I',I'')$ denotes the triple of
independent links emanating from a given vertex, we can write the
volume as $\hat{V}_{\rm v}= \sqrt{|\hat{F}_{{\rm v},I} 
\hat{F}_{{\rm v},I'}\hat{F}_{{\rm
v},I''}|}$. Thus,
\begin{equation}
\hat{B}_{{\rm v},I}= \frac{1}{4\pi \gamma G\hbar}
\left(\hat{\eta}_{{\rm v},I}^{\dagger}\sqrt{|\hat{F}_{{\rm v},I} \hat{F}_{{\rm v},I'}\hat{F}_{{\rm
v},I''}|} \hat{\eta}_{{\rm v},I}-
\hat{\eta}_{{\rm v},I}\sqrt{|\hat{F}_{{\rm v},I} \hat{F}_{{\rm v},I'}\hat{F}_{{\rm
v},I''}|} \hat{\eta}_{{\rm v},I}^{\dagger}\right)\,.
\end{equation}
As in the general representation, the basic operators $\hat{F}_{{\rm
v},I}$ and $\hat{\eta}_{{\rm v},I}$ satisfy the commutator identity
\Eq{core} while $\hat{\eta}_{{\rm v},I}$ commutes with $\hat{F}_{{\rm
v},I'}$ and $\hat{F}_{{\rm v},I''}$. Moreover, $\hat{\eta}_{{\rm
v},I}$ satisfies the reality condition $\hat{\eta}_{{\rm
v},I}\hat{\eta}_{{\rm v},I}^{\dagger}=1$. It turns out that these
identities are sufficient to derive the form of inverse-triad
corrections in a semiclassical expansion, irrespective of what state
is used beyond general requirements of semiclassicality.

We consider the two operators $\hat{\eta}_{{\rm v},I}|\hat{F}_{{\rm
v},I}|^{1/2}\hat{\eta}_{{\rm v},I}^{\dagger}$ and $\hat{\eta}_{{\rm
v},I}^{\dagger}|\hat{F}_{{\rm v},I}|^{1/2}\hat{\eta}_{{\rm v},I}$
separately.\footnote{The factors of $\sqrt{|\hat{F}_{{\rm v},I'}|}$
and $\sqrt{|\hat{F}_{{\rm v},I''}|}$ quantize positive powers of the
densitized triad and do not give rise to inverse-volume corrections.}
They can be simplified using the basic commutators \Eq{core} and
$[\hat{\eta}_{{\rm v},I}^{\dagger},\hat{F}_{{\rm
v},I}]=4\pi\gamma\ell_{\rm Pl}^2\hat{\eta}_{{\rm v},I}^{\dagger}$. We
can thus reorder terms so as to bring $\hat{\eta}_{{\rm v},I}$ right
next to $\hat{\eta}_{{\rm v},I}^{\dagger}$, and then cancel them using
the reality condition.  Reordering according to $\hat{\eta}_{{\rm
v},I}\hat{F}_{{\rm v},I}=(\hat{F}_{{\rm v},I}-4\pi\gamma\ell_{\rm
Pl}^2)\hat{\eta}_{{\rm v},I}$ and $\hat{\eta}_{{\rm
v},I}^{\dagger}\hat{F}_{{\rm v},I}=(\hat{F}_{{\rm
v},I}+4\pi\gamma\ell_{\rm Pl}^2)\hat{\eta}_{{\rm v},I}^{\dagger}$
leads to
\begin{equation}
 \hat{\eta}_{{\rm v},I}|\hat{F}_{{\rm v},I}|^{1/2}\hat{\eta}_{{\rm
     v},I}^{\dagger}= |\hat{F}_{{\rm v},I}-4\pi\gamma\ell_{\rm
   Pl}^2|^{1/2}\,, \qquad 
\hat{\eta}_{{\rm v},I}^{\dagger}|\hat{F}_{{\rm v},I}|^{1/2}
 \hat{\eta}_{{\rm v},I}= |\hat{F}_{{\rm v},I}+4\pi\gamma\ell_{\rm Pl}^2|^{1/2}\,.
\end{equation}
In the classical limit $\hbar\to0$, these expressions in
$\hat{B}_{{\rm v},I}$ result in a derivative by $F_{{\rm v},I}$, as
required by the Poisson bracket relationship
(\ref{cotriadPoisson}). For inverse-volume effects we are interested
in the leading quantum corrections with $\hbar\neq0$, which arise in
different forms. First, because the operator is nonlinear in the basic
ones $\hat{F}_{{\rm v},I}$ and $\hat{\eta}_{{\rm v},I}$, classical
expressions will be corrected by terms involving the moments of a
state: As always in quantum physics, the expectation value
$\langle\hat{B}_{{\rm v},I}\rangle$ does not have the classically
expected relationship with expectation values of the basic
operators. We can compute these corrections by following the
principles of canonical effective dynamics, substituting
$\langle\hat{F}_{{\rm v},I}\rangle+ (\hat{F}_{{\rm
v},I}-\langle\hat{F}_{{\rm v},I}\rangle)$ for $\hat{F}_{{\rm v},I}$
and performing a formal expansion by $\hat{F}_{{\rm
v},I}-\langle\hat{F}_{{\rm v},I}\rangle$:
\begin{eqnarray} \label{OpExpInv}
 \hat{\eta}_{{\rm v},I}|\hat{F}_{{\rm v},I}|^{1/2}\hat{\eta}_{{\rm v},I}^{\dagger}&=& 
|\hat{F}_{{\rm v},I}-4\pi\gamma\ell_{\rm Pl}^2|^{1/2}\nonumber\\
&=&
 |\langle\hat{F}_{{\rm v},I}\rangle-4\pi\gamma\ell_{\rm Pl}^2|^{1/2} \sum_{k=0}^{\infty} 
 \binom{1/2}{k}
 \frac{(\hat{F}_{{\rm v},I}- \langle\hat{F}_{{\rm v},I}\rangle)^k}{|\langle\hat{F}_{{\rm v},I}\rangle-
   4\pi\gamma\ell_{\rm Pl}^2|^k}\,,\\
 \hat{\eta}_{{\rm v},I}^{\dagger}|\hat{F}_{{\rm v},I}|^{1/2}\hat{\eta}_{{\rm v},I}&=& 
|\hat{F}_{{\rm v},I}+4\pi\gamma\ell_{\rm Pl}^2|^{1/2}\nonumber\\
&=&
 |\langle\hat{F}_{{\rm v},I}\rangle+4\pi\gamma\ell_{\rm Pl}^2|^{1/2} \sum_{k=0}^{\infty} 
\binom{1/2}{k}
 \frac{(\hat{F}_{{\rm v},I}- \langle\hat{F}_{{\rm v},I}\rangle)^k}{|\langle\hat{F}_{{\rm v},I}\rangle+
   4\pi\gamma\ell_{\rm Pl}^2|^k}\,.
\end{eqnarray}
(This expansion can be made well-defined and analyzed in the context
of Poisson geometry of algebraic state spaces
\cite{EffAc,EffCons,EffConsRel}.)

All terms in $\hat{F}_{{\rm v},I}-\langle\hat{F}_{{\rm v},I}\rangle$
will either vanish upon taking an expectation value ($k=1$ in the
expansion) or give rise to moments of the quantum state used to
compute the expectation value ($k\geq 2$, with fluctuations and
correlations arising for $k=2$). The precise values of the moments and
their dynamics depend on the state used and in fact encode the state
dependence of the theory, but for a semiclassical state they satisfy,
by definition, the hierarchy $\langle(\hat{F}_{{\rm
v},I}-\langle\hat{F}_{{\rm v},I}\rangle)^k\rangle\sim \hbar^{k/2}$. This
notion of semiclassicality is a very general one; it does not require
us to choose a particular shape of the state, such as a Gaussian.

The moment terms imply an important form of quantum
corrections in the context of quantum back-reaction. Such corrections
arise from different sources in the Hamiltonians, which will all have
to be combined and analyzed. We will not enter such an analysis here,
but rather note that even if we disregard quantum back-reaction,
quantum corrections do remain: we have 
\ba
&& \frac{1}{4\pi\gamma G\hbar} (\hat{\eta}_{{\rm v},I}^{\dagger}|\hat{F}_{{\rm v},I}|^{1/2}
 \hat{\eta}_{{\rm v},I}- 
 \hat{\eta}_{{\rm v},I}|\hat{F}_{{\rm v},I}|^{1/2}\hat{\eta}_{{\rm v},I}^{\dagger})
 \nonumber\\
&&\qquad=\frac{|\langle\hat{F}_{{\rm v},I}\rangle+4\pi\gamma\ell_{\rm Pl}^2|^{1/2}-
   |\langle\hat{F}_{{\rm v},I}\rangle-
4\pi\gamma\ell_{\rm Pl}^2|^{1/2}}{4\pi\gamma\ell_{\rm Pl}^2}+\cdots\,, \label{InvTriadOp}
\ea
where the dots indicate moment terms dropped. This expression includes
inverse-volume corrections, computed for general semiclassical
states. It matches with expressions derived directly from triad
eigenstates \cite{InvScale,QuantCorrPert}, which are not semiclassical
but, as proven here, provide reliable information about inverse-volume
corrections. More general semiclassical states do not introduce
additional dependence of inverse-volume corrections on connection
components or curvature, they just introduce moment terms which
contribute to quantum back-reaction. (Such an extra dependence may
arise from non-Abelian properties of the theory \cite{DegFull}, which
are not strong for perturbative inhomogeneities.)

\subsection{Correction functions}\label{23}

Corrections to classical Hamiltonians in which inverse triad
components appear can be captured by introducing correction
functions such as
\begin{equation}
 \alpha(a):=\frac{|L(a)^2+4\pi\gamma\ell_{\rm Pl}^2|^{1/2}-
   |L(a)^2-
4\pi\gamma\ell_{\rm Pl}^2|^{1/2}}{4\pi\gamma\ell_{\rm Pl}^2}\, L(a)\,,
\end{equation}
obtained by identifying 
\be
\langle\hat{F}_{{\rm v},I}\rangle= L^2(a)
\ee
with the discreteness scale (depending on the scale factor in the presence of
lattice refinement \cite{InflObs}). The multiplication of inverse-volume corrections
by $L(a)$ ensures that $\alpha(a)\sim 1$ in the classical limit, but
strong corrections can arise for small $L$. Our derivations apply to
small deviations from the classical value, for which consistent
implementations in the dynamics are available. We can thus expand
\begin{equation}
 \alpha(a)= 1+\alpha_0 \dpl+\cdots\,,
\end{equation}
with $\dpl:= (\ell_{\rm Pl}/L)^m$ for $m=4$ in the above derivation, and
  the dots indicating powers higher than $m$.

For $\langle\hat{F}_{{\rm v},I}\rangle\gg\ell_{\rm Pl}^2$ inverse-volume
corrections become very small, but they are significant if
$\langle\hat{F}_{{\rm v},I}\rangle$ is about as large as a Planck area or
smaller. Bringing in our discreteness scale, leading inverse-volume
corrections can be expressed in terms of the quantity
\be
\dpl= \left(\frac{\ell_{\rm Pl}}{L}\right)^4= \left(\frac{\ell_{\rm Pl}^3}{v}\right)^{\frac{4}{3}}
\ee
(using $m=4$ from now on). If $L$ or $v$ is constant, $\dpl$ is
constant and inverse-volume corrections merely amount to rescaling
some expressions in Hamiltonians. More generally, however, the
dynamical nature of a discrete state suggests that $L$ and $v$ change
in time or, in cosmology, with respect to the scale factor $a$. We
parameterize this dependence as
\begin{equation} \label{dpl2}
\dpl\propto a^{-\sigma}
\end{equation} 
with $\sigma\geq 0$; see \cite{InflObs} for a discussion of possible
values of $\sigma$ and its relation to quantization parameters.

In order to compare inverse-volume with holonomy corrections, we write
\begin{equation}
 \dpl= \left(\frac{8\pi G}{3}\rho_\textsc{qg}
 \ell_{\rm Pl}^2\right)^2=
 \left(\frac{8\pi}{3}\frac{\rho_\textsc{qg}}{\rho_{\rm
 Pl}}\right)^2= \left(\frac{8\pi}{3} \frac{\rho}{\rho_{\rm
 Pl}} \delta_{\rm hol}^{-1}\right)^2\,.
 \label{dpl}
\end{equation}
The second equality shows that inverse-volume corrections are 
considerable and of the order one when the quantum-gravity density is
close to the Planck density. Inverse-volume corrections thus behave
very differently from what is normally expected for quantum gravity,
where the Planck density is often presupposed as the quantum-gravity
scale. In loop quantum gravity, this scale must be sufficiently small
compared to the Planck density in order to be consistent with
inverse-volume corrections.

The last expression in eq.~(\ref{dpl}) is useful in order to compare
holonomy with inverse-volume corrections. Inverse-volume corrections
are usually suppressed by a factor of $\rho/\rho_{\rm Pl}$, as
expected for quantum-gravity effects, but there is an extra factor of
$\delta_{\rm hol}^{-1}$. For small densities, holonomy corrections are
small, but inverse-volume corrections may still be large because they
are magnified by the inverse of $\delta_{\rm hol}$. As the energy
density decreases in an expanding universe, holonomy corrections fall
to small values, and in this way begin to magnify inverse-volume
corrections. For instance, in an inflationary regime with a
typical energy scale of $\rho\sim 10^{-10}\rho_{\rm Pl}$, we can use
(\ref{dpl}) to write $\delta_{\rm hol} \sim
10^{-9}/\sqrt{\dpl}$. Having small holonomy corrections of size
$\delta_{\rm hol}<10^{-6}$ then requires inverse-volume correction
larger than $\dpl>10^{-6}$.  This interplay of holonomy and
inverse-volume corrections makes loop quantum gravity testable because
it leaves only a finite window for consistent parameter values, rather
than just providing Planckian upper bounds. It also shows that
inverse-volume corrections become dominant for sufficiently small
densities, as they are realized even in high-energy scenarios of
inflation.

In this context, it is worthwhile to comment on a comparison of the
corrections derived here, assuming a nearly isotropic but explicitly
inhomogeneous discrete state, with their form in pure minisuperspace
quantizations. In inverse-volume as well as holonomy corrections, we
referred to elementary building blocks of a discrete state, the
plaquette areas in inverse-volume corrections and edge lengths in
holonomy corrections. A pure minisuperspace quantization would
primarily make use of macroscopic parameters such as the volume of
some region (or the scale factor). The number of discrete blocks, such
as ${\cal N}$ introduced above, is not available, and thus it is more
difficult to refer to local microscopic quantities such as $F_{{\rm
v},I}$.

For curvature or the Hubble parameter, local quantities are easier to
introduce and to use in holonomy corrections, but inverse-volume
expressions must refer to quantities of size, which cannot be
expressed microscopically in a pure minisuperspace context. As a
consequence, inverse-volume corrections have often been misrepresented
in loop quantum cosmology. Without referring to ${\cal N}$, as it is
introduced in the lattice-refinement formulation of loop quantum
cosmology, one can only use the macroscopic volume of some region
instead of the microscopic $F_{{\rm v},I}$.\footnote{As mentioned
earlier, in the inhomogeneous theory we can use the full volume or the
size of any region in inverse-volume corrections because most
plaquette contributions, which do not intersect the edge of the
holonomies used, drop out.  In homogeneous models, on the other hand,
all plaquettes are equivalent and correspond to the same degree of
freedom. The choices must thus be specified carefully in order to
avoid minisuperspace artifacts.} Inverse-volume corrections become
smaller for larger $F_{{\rm v},I}$, and thus substituting this
quantity by a macroscopic size suppresses the corrections. Any such
suppression is merely an artifact of using the wrong expressions for
the corrections based solely on minisuperspace considerations. Using a
macroscopic volume also makes the corrections dependent on the size of
the chosen region, which is another artificial dependence on extra
parameters; because of this, LQC inverse-volume corrections have been
often interpreted as problematic or even unphysical. The derivation
shown here solves these problems; see also the following subsection.

As already seen, inverse-volume corrections show unexpected properties
in terms of their dependence on the density, and regimes in which they
are strong. Another unexpected property is seen in their influence on
spacetime structure, with important consequences for cosmological
perturbation theory. Inverse-volume corrections are not just of
higher-curvature type in an effective action, but they deform the
usual gauge algebra of generally covariant systems, generating
spacetime diffeomorphisms. This deformation, as discussed in more
detail in the following calculations, leads to characteristic
cosmological effects. In a conceptual context, moreover, it allows us
to distinguish inverse-volume corrections from the other types
encountered in loop quantum gravity: holonomy corrections and 
quantum back-reaction. 

A closer look at the algebra of constraints generating
the gauge transformations reveals that deformations of the algebra
introduced by inverse-volume corrections cannot be undone by including
holonomy corrections or quantum back-reaction \cite{InflObs}. Holonomy corrections
imply higher-order terms in the constraints depending on the
connection nonpolynomially, or at least on the background connection
if an expansion by inhomogeneities is done. No such terms arise for
inverse-volume corrections, and no cancellation is possible. Quantum
back-reaction, on the other hand, comes from terms including moments
of a state, as alluded to in our derivation of inverse-volume
corrections. The dependence on the moments remains if one computes the
constraint algebra, in such a way that corrections from quantum
back-reaction cannot cancel deformations implied by inverse-volume
corrections, either. Since the characteristic effects analyzed here
are a consequence of nontrivial deformations of the algebra, we can
safely conclude that including only inverse-volume corrections does
give a reliable picture, because they cannot be cancelled by the
other, more complicated corrections. Of course, it remains of interest to study the inclusion of other effects such as the curvature of the universe, and the simultaneous competition between inverse-volume and other quantum corrections in a more complete dynamical analysis.

\subsection{Consistency}\label{24}

Most of the properties and consequences of inverse-volume
corrections are unexpected and unfamiliar. It is then perhaps not
surprising that there are at least four main objections to the
physical significance of effective LQC dynamics with inverse-volume
corrections, which are popularly encountered in the literature and in
scientific debates. It is claimed that (i) these corrections are
ill-defined in a pure minisuperspace context and a flat universe, (ii)
no rigorous derivation in the more involved inhomogeneous context
(taking into account lattice refinement) has been provided so far,
(iii) even if a derivation were possible, the inflationary energy
scale would be too low for volume/curvature corrections to be sizable,
and (iv) even setting aside the issue of their size, the analysis
would remain incomplete because we do not know how these corrections
compete with holonomy modifications of the dynamics. As an example for
the claimed incompleteness of correction functions used, the
independence of inverse-volume corrections of the connection or
curvature has been criticized as physically unjustified.

We had already partially answered some of these objections elsewhere
\cite{InflObs}. First, let us summarize the main arguments advanced
there:
\begin{enumerate}
\item[(i)] In a realistic cosmological scenario, there is no conformal
invariance of the scale factor and the correct way to implement the
quantum dynamics is to consider the natural cell subdivision of space
and how these cells evolve in time: this is the lattice refinement
picture. In this perspective, interpretational difficulties regarding
quantum corrections appear to be just an artifact of the idealized
homogeneous and isotropic setting of pure minisuperspace models. 
\item[(ii)] Although a rigorous derivation is desirable, the
motivations of lattice refinement are natural in the perspective of
the full quantum theory and there is no conceptual obstacle in
relaxing the parametrization obtained in a pure
minisuperspace.\footnote{Sometimes, the argument is advanced that the
minisuperspace parametrization (in particular, the so-called improved
dynamics) is the only one producing a constant critical density and a
robust bounce picture. This argument is invalid for two reasons. On
one hand, even the improved dynamics parametrization does \emph{not}
give a constant critical density unless quantization ambiguities are
tuned to certain specific values \cite{CH}; the time-dependent
modification comes from inverse-volume corrections in the
gravitational sector, which are nonzero in general. On the other hand,
within the lattice parametrization a constant critical density, if
desired, can be obtained, indeed.}  Moreover, one cannot simply
suppress inverse-volume corrections by a regularization procedure, as
occasionally suggested by taking the limit of ${\cal V}_0\to\infty$ in
cases where these corrections are ${\cal
V}_0$-dependent. Inverse-volume corrections do appear in the full
quantum theory and play an important role for well-defined
Hamiltonians. If they disappeared by a regularization procedure in
minisuperspace models, one should explain why they are absent in a
cosmological setting but not otherwise. Furthermore, there is tension
between the requirement of closure of the inhomogeneous constraint
algebra and the minisuperspace parametrization \cite{InflObs}, which
demands clarifications; although the lattice parametrization is so far
implemented semi-heuristically in calculations of effective constraint
algebras, it does accommodate anomaly cancellation.
\item[(iii)] Since the gauge symmetry of the model is
deformed by quantum corrections, the very structure of spacetime is
modified locally but everywhere; thus, one expects effects larger than
in traditional scenarios of standard general relativity with
higher-order curvature terms. In \cite{InflObs} we found qualitative
theoretical estimates of these effects which are several orders of
magnitude larger than minisuperspace estimates (and, interestingly,
rather close to experimental bounds \cite{BCT} in terms of orders of
magnitude). However, the lack of control over the putative quantum
gravity characteristic scale (hidden in the quantum corrections) makes
it difficult to assess its importance within inflation. 
\item[(iv)] We argued
that other quantum corrections would not cancel inverse-volume effect
because of the radically different way in which they affect the
dynamics. Of course, the issue of comparing inverse-volume and
holonomy corrections remains of interest for the community, but one
does not expect that miraculous cancellations happen 
between the two.
\end{enumerate}

The results of the present section serve to further address the above
objections and provide final clarifications for several of them. For
the first time, we have embedded inverse-volume corrections in
inhomogeneous models, using the lattice refinement picture and working
at the kinematical level, thus giving fresh insight to these
issues. In particular: 
\begin{enumerate}
\item[(i)-(ii)] When the phase space volume is
associated with an individual homogeneous cell rather than a fiducial
volume (as done in pure minisuperspace), the lattice parametrization
emerges naturally and a quantum-gravity scale replaces unphysical
quantities in inverse-volume corrections. Correction functions are
completely independent of comoving volumes such as ${\cal V}_0$ and
there is no regularization needed to make them disappear. 
\item[(iii)] Surprisingly, the magnitude of these corrections can be argued to be
large at mesoscopic scales, even when densities are far away from
Planckian values such as during inflation. In cosmological models,
quantum corrections are relevant not just near a bounce at Planckian
density. 
\item[(iv)] The basic noncancellation between inverse-volume
effects and other, presently uncontrolled quantum corrections is
reiterated with novel arguments. Modifications of the classical
constraint algebra by inverse-volume corrections cannot cancel with
those from holonomy corrections, nor with terms from quantum
back-reaction. Holonomy corrections provide an additional connection
dependence of almost-periodic type in the constraints, while
inverse-volume corrections as shown here have only weak connection
dependence. Inverse-triad corrections are also independent of moments
of a state, as they would determine quantum back-reaction. The
structure of the Poisson algebra on the quantum phase space, including
expectation values and moments, shows that neither the connection-dependent terms
of the form of holonomy corrections nor moment terms describing
quantum back-reaction can cancel the terms of inverse-volume
corrections. If the constraint algebra is modified by inverse-volume
corrections, it must remain modified when all corrections are
included. Thus, also the presence of effects larger than usually
expected in quantum gravity is general.
\end{enumerate}
To summarize, loop quantum cosmology implies the presence of
inverse-volume corrections in its cosmological perturbation equations.
In their general parametrization, the corrections depend only on
triad variables simply because they depend on a quantum scale whose
\emph{dynamical} nature is encoded by the background scale
factor. This conclusion is a result of the derivations presented here,
not an assumption. Also, their power-law form as a function of the
scale factor is suggested by very general semiclassical considerations which do
not further restrict the class of states.

\section{Inflationary observables}\label{ifos}

With a consistent implementation of inverse-volume corrections at
hand, a complete set of cosmological perturbation equations
follows. These equations have been derived 
elsewhere \cite{BH1}-\cite{BHKS2}, starting with a
constraint analysis. Here we continue to prepare these equations for a
convenient cosmological investigation, which we then exploit to find
observational bounds on some parameters.

The slow roll approximation is assumed precisely for the same reasons as in standard inflation: it is an {\it Ansatz}, it is the definition itself of inflation. In LQC there exists also a super-inflationary regime where the universe super-accelerates because of purely geometric effects, and the background attractor is not de Sitter. However, in that regime the constraint algebra has not been shown to close, and we have no rigorous control over the ensuing physics. Here, we simply assume that (i) inflation takes place thanks to a scalar field slowly rolling down its potential, and (ii) that this happens completely in the large-volume regime, where quantum corrections are small (and the algebra closes).

\subsection{Hubble slow-roll tower}\label{hsr}

The slow-roll parameters as functions of the Hubble rate are defined 
starting from the background equations of motion, which also determine 
the coefficients of the linear perturbation equations. 
In the presence of inverse-volume corrections, the effective Friedmann and Klein--Gordon equations read 
\be\label{frw}
\cH^2=\frac{\k^2}{3}\,\a\left[\frac{{\vp'}^2}{2\nu}+pV(\vp)\right]
\ee 
and 
\be\label{kg} 
\vp''+2\cH\left(1-\frac{\rmd\ln\nu}{\rmd\ln
p}\right)\vp'+\nu p V_{,\vp}=0\,, 
\ee 
respectively, where primes denote derivatives with respect to 
conformal time $\tau:=\int \rmd t/a$, $\cH:= a'/a=aH$, 
$\kappa^2=8\pi G$, $p=a^2$, $G$ is Newton's constant, and $\vp$ is a real 
scalar field with potential $V(\vp)$. 
Following section \ref{invo}, the LQC correction functions are of the form
\ba
\a &=& 1+\a_0\dpl,\label{an}\\ 
\nu &=& 1+\nu_0\dpl,\label{cn}
\ea
where $\a_0$ and $\nu_0$ are constants and 
\be
\dpl\propto a^{-\s}
\ee
is a quantum correction (\ref{dpl2}) whose time dependence is
modelled as a power of the scale factor (here $\s>0$ is another
constant). The proportionality factor will never enter the analysis
explicitly but, in the derivation of the perturbation equations,
it is assumed that $\dpl< 1$. Consistently, throughout the paper we
use the equality symbol $=$ for expressions valid up to $O(\dpl)$
terms, while we employ $\approx$ for relations where the slow-roll
approximation has been used.  The latter holds when the following
slow-roll parameters are small:
\ba
\e &:=& 1-\frac{\cH'}{\cH^2}\nonumber\\
 &=& \frac{\k^2}{2}\frac{\vp'^2}{\cH^2}\left\{1+\left[\a_0
 +\nu_0\left(\frac{\s}{6}-1\right)\right]\dpl\right\}+\frac{\s\a_0}{2}\dpl\,,\label{edp}\\
\eta &:=& 1-\frac{\vp''}{\cH\vp'}\,.
\ea
The conformal-time derivatives of $\epsilon$ and $\eta$ are
\ba
\e' &=& 2\cH\epsilon(\epsilon-\eta)-\s\cH\tilde{\epsilon} \dpl\,,\\
\eta' &=& \cH(\epsilon\eta-\xi^2)\,,
\ea
where
\be\label{vare}
\tilde{\epsilon}:= \a_0\left(\frac{\s}{2}+2\e-\eta\right)
+\nu_0\left(\frac{\s}6-1\right)\e\,. 
\ee

The inflationary spectra were computed in \cite{InflObs}. 
The scalar power spectrum is 
\be
\label{powerscalar}
{\cal P}_{\rm s}=\frac{G{\cal H}^2}{\pi a^2 \epsilon}
(1+\gamma_{\rm s} \dpl)\,,
\ee
where 
\be
\gamma_{\rm s} := \nu_0 \left( \frac{\sigma}{6}+
1 \right)+\frac{\sigma \alpha_0}{2\epsilon}
-\frac{\chi}{\sigma+1}\,,\qquad
\chi := \frac{\sigma \nu_0}{3} \left(\frac{\sigma}{6}+1 \right)+
\frac{\alpha_0}{2} \left( 5-\frac{\sigma}{3} \right)\,.
\ee
Equation \Eq{powerscalar} is evaluated at the time $k=\cH$ when the 
perturbation with comoving wavenumber $k$ crosses the Hubble horizon. 
Using the fact that
\be\label{dDd}
\dpl'=-\s\cH\dpl
\ee
and $'\approx\cH\, d/d\ln k$, 
the scalar spectral index $n_{\rm s}-1:= \rmd\ln {\cal P}_{\rm s}/\rmd\ln k$ reads
\be
n_{\rm s}-1 = 2\eta-4\e+\s\g_{n_{\rm s}}\dpl\,,\label{ns}
\ee
where
\be
\g_{n_{\rm s}} := \frac{\tilde{\epsilon}}{\e}
-\a_0\left(1-\frac{\eta}{\e}\right)-\g_{\rm s} 
= \a_0-2\nu_0+\frac{\chi}{\s+1}\,,
\ee
while the running $\a_{\rm s}:= \rmd n_{\rm s}/\rmd\ln k$ is
\be
\a_{\rm s}=2(5\e\eta-4\e^2-\xi^2)+
\s(4\tilde{\epsilon}-\s \g_{n_{\rm s}})\dpl\,.
\label{als}
\ee
This shows that, for $\sigma=O(1)$, the running can be 
as large as $\dpl$.
In this case, the terms higher than the running can give rise 
to the contribution of the order of $\dpl$.
In section \ref{ksr1} we shall address this issue properly.

The tensor power spectrum is 
\be
\label{powertensor}
{\cal P}_{\rm t}=\frac{16G{\cal H}^2}{\pi a^2}
(1+\gamma_{\rm t} \dpl)\,,
\qquad
\gamma_{\rm t}=\frac{\sigma-1}{\sigma+1}\alpha_0\,,
\ee
while its index $n_{\rm t}:= \rmd\ln {\cal P}_{\rm t}/\rmd\ln k$ and 
running $\a_{\rm t}:= \rmd n_{\rm t}/\rmd\ln k$ are, respectively,
\be
n_{\rm t}= -2\e-\s\g_{\rm t}\dpl\,,
\label{nt}
\ee
and
\be
\alpha_{\rm t}=-4\epsilon (\epsilon-\eta)+
\sigma (2\tilde{\epsilon}+\sigma \gamma_{\rm t} )\dpl\,.
\label{alt}
\ee
The tensor-to-scalar ratio $r:={\cal P}_{\rm t}/{\cal P}_{\rm s}$ 
combines with the tensor index into a consistency relation:
\ba
\label{tts2}
r &=&16 \epsilon \left[1+(\gamma_{\rm t}-\gamma_{\rm s} )\dpl
\right] \nonumber \\
&=& -8\{n_{\rm t}+[n_{\rm t}(\g_{\rm t}-\g_{\rm s})+\s\g_{\rm t}]\dpl\}\,.
\ea
When $\dpl=0$, all the above formulas agree with the standard 
classical scenario \cite{review}.

\subsection{Potential slow-roll tower}\label{vsr}

To constrain the inflationary potential against observations, it is convenient 
to recast the cosmological observables in terms of the tower of slow-roll parameters 
written as functions of $V$ and its derivatives. From eqs.~\Eq{frw} and \Eq{edp} 
we have
\ba
\frac{\k^2}{2}\frac{\vp'^2}{\cH^2} &=&  \e-\left\{\frac{\s\a_0}{2}+\e\left[\a_0+\nu_0\left(\frac{\s}{6}-1\right)\right]\right\}\dpl\,,\label{edp2}\\
\nu pV &=& \left(3\frac{\nu}{\a}-\frac{\k^2}{2}\frac{{\vp'}^2}{\cH^2}\right)\frac{\cH^2}{\k^2}\nonumber\\
       &=& \frac{3 \cH^2}{\k^2} \left(1-\frac{\e}{3}+\left\{\nu_0+\a_0\left(\frac{\s}{6}-1\right)+\frac{\e}{3}\left[\a_0+\nu_0\left(\frac{\s}{6}-1\right)\right]\right\}\dpl\right),\\
\label{dV1}
V_{,\vp}  &=& -\frac{3\cH \vp'}{\nu p}\left(1-\frac{\eta}{3}+\frac{\s\nu_0}{3}\dpl\right)\,,\\
\label{dV2}
V_{,\vp\vp} &=& \frac{\cH^2}{\nu p}\left[3(\epsilon+\eta)-\eta^2-\xi^2
-(3-\s-\epsilon-2\eta)\s\nu_0\dpl\right]\,,\\
\label{dV3}
V_{,\vp \vp \vp} &=& -\frac{3\cH^3}{\nu p \vp'}
\left(\xi^2+3\epsilon \eta-\s\left\{\nu_0 \left[\left(1-\frac{2\s}{3}\right)\eta+(3-\s)\left(\e
+\frac{\s}{3}\right)\right.\right.\right.\nonumber\\
&&\qquad\qquad\qquad\qquad\qquad\qquad\left.\left.\left.-\frac43\e\eta-\frac13\eta^2-\xi^2\right]-
\tilde{\epsilon} \right\} \dpl \right)\,.
\ea
The first three elements of the tower are
\be
\e_\V:=\frac{1}{2\k^2}\left(\frac{V_{,\vp}}{V}\right)^2\,,
\qquad 
\eta_\V:=\frac{1}{\k^2}\frac{V_{,\vp\vp}}{V}\,,
\qquad
\xi_\V^2 := \frac{V_{,\vp}V_{,\vp \vp \vp}}{\kappa^4 V^2}\,.
\ee
On using eqs.~(\ref{dV1})--(\ref{dV3}), we have the following technical expressions:
\bs\ba
\e_\V &\approx&\e+\left\{-\frac{\s\a_0}{2}+\frac{\s\a_0}{3}\eta
+\left[\a_0\left(1-\frac{2\s}{3}\right)
+\nu_0\left(\frac{\s}{2}-1\right)\right]\e\right\}\dpl\,,\\
\eta_\V &\approx&\e+\eta+\left\{\s\nu_0\left(\frac{\s}{3}-1\right)
+\left[\a_0\left(1-\frac{\s}{6}\right)+\nu_0\left(\frac{2\s}{3}-1\right)
\right]\eta\right.\nonumber\\
&&\qquad\qquad\left.+\left[\a_0\left(1-\frac{\s}{6}\right)
+\nu_0\left(\frac{\s^2}{9}-1\right)\right]\e\right\}\dpl\,,\\
\xi_{\V}^2 &\approx& \xi^2+3\e\eta+[\s f_\xi(\e,\eta)+g_\xi(\e,\eta,\xi^2)]\dpl\,,\\
f_\xi(\e,\eta) & := & \sigma \left[ \frac{\alpha_0}{2}+\nu_0 \left( \frac{\sigma}{3}
-1 \right) \right]
+\left[2\a_0\left(\frac{\s}{6}+1\right)-\nu_0\left(4-\frac{\s}{2}-\frac{2\s^2}{9}\right)
\right]\e\nonumber\\
&&-\left[\a_0\left(\frac{\s}{6}+1\right)+\nu_0
\left(1-\s+\frac{\s^2}{9}\right)\right]\eta\,,\nonumber\\
g_\xi(\e,\eta,\xi^2) & := & 
\left[\a_0\left(4+\frac{\s}{2}\right)-\nu_0\left(8-\frac{4\s}{3}-\frac{\s^2}{3}\right) 
\right]\frac{\s}{3}\,\e^2
+\left[\a_0+2\nu_0\left(1-\frac{\s}{3}\right)\right]\frac{\s}{3}\eta^2\nonumber\\
&&+\left[\a_0\left(6-\frac{7\s}{3}-\frac{\s^2}{9}\right)-\nu_0\left(6-3\s
-\frac{5\s^2}{18}+\frac{2\s^3}{27}\right)\right]\e\eta\nonumber\\
&&
+2\left[\a_0\left(1-\frac{\s}{6}\right)+\nu_0\left(\frac{2\s}{3}-1\right)\right]\xi^2\,,
\ea\es
which give the inversion formulas
\bs\ba
\e &\approx&\e_\V+ \left\{\frac{\s\a_0}{2}-\left[\a_0\left(1-\s\right)
+\nu_0\left(\frac{\s}{2}-1\right)\right]\e_\V 
-\frac{\s\a_0}{3}\eta_\V \right\} \dpl\,,\\
\eta &\approx&\eta_\V-\e_\V \nonumber \\
& &-\left\{
\s\left(\frac{\alpha_0}{2}+\frac{\sigma \nu_0}{3}-\nu_0 \right)
+\left[\a_0\left(\s-1\right)+\nu_0\left(1-\frac{7\s}{6}+\frac{\s^2}{9}\right)\right]\e_\V\right.\nonumber\\
&&\qquad
\left.+\left[\a_0\left(1-\frac{\s}{2}\right)+\nu_0\left(\frac{2\s}{3}-1\right)\right]\eta_\V \right\}\dpl\,,\\
\xi^2 &\approx& \xi^2_\V+3\e_\V^2-3\e_\V\eta_\V \nonumber\\
&&+\biggl\{\s^2 \left[\nu_0\left(1-\frac{\s}{3}\right)-\frac{\a_0}{2}\right]
+\s^2\left[-\frac{\a_0}{2}+\nu_0\left(\frac32-\frac{\s}{3}\right)\right]\e_\V\nonumber\\
&&+\s\left[\frac{\a_0}{2}\left(\frac{\s}{3}-1\right)+\nu_0\left(1-\s+\frac{\s^2}{9}\right)\right]\eta_\V+\frac23\left[\a_0+\nu_0\left(\frac{\s}{3}-1\right)\right]\s\eta_\V^2 \nonumber \\
& &+\left[-\a_0\left(6-3\s+\frac{5\s^2}{18}\right)+\nu_0\left(6-4\s
+\frac{7\s^2}{18}-\frac{5\s^3}{27}\right)\right]\e_\V^2\nonumber \\
&&+\left[\a_0\left(6-\frac{7\s}{2}+\frac{\s^2}{9}\right)-\nu_0\left(6-\frac{35\s}{6}+\frac{13\s^2}{18}-\frac{2\s^3}{27}\right)\right]\e_\V\eta_\V \nonumber \\
& &+\left[\a_0\left(\frac{\s}{3}-2\right)+2\nu_0\left(1-\frac{2\s}{3}\right)\right]\xi_\V^2
\biggr\} \dpl\,.
\ea\es

We can now rewrite the cosmological observables. The scalar index \Eq{ns} and its running \Eq{als} become
\ba
n_{\rm s}-1 &=& -6\e_\V+2\eta_\V
-c_{n_{\rm s}}\dpl\,,\label{ns2} \\
\alpha_{\rm s} &=& -24 \e_\V^2+16\e_\V\eta_\V
-2\xi_\V^2+c_{\a_{\rm s}}\dpl\,,\label{alphas}
\ea
where
\bs\ba
\label{gammas}
c_{n_{\rm s}} &=& f_{\rm s}-\left[6\a_0\left(1-\s\right)-\nu_0\left(6-\frac{13\s}{3}+\frac{2\s^2}{9}\right)\right]\e_\V\nonumber\\
&&-\left[\a_0\left(\frac{7\s}{3}-2\right)+2\nu_0\left(1-\frac{2\s}{3}\right)\right]\eta_\V\,,
\ea
\ba
c_{\a_{\rm s}} &=& 
\s f_{\rm s} +\left[\a_0\s(\s-6)+\nu_0\s\left(6-\frac{17\s}{3}+\frac{2\s^2}3\right)\right] \e_\V\nonumber\\
&&+
\left[\a_0\s\left(2-\frac{\s}3\right)-2\nu_0\s\left(1-\s+\frac{\s^2}{9}\right)\right] \eta_\V \nonumber \\
& & +\left[\a_0\left(48-42\s+\frac{5\s^2}{9}\right)-\nu_0\left(48-\frac{98\s}{3}+\frac{17\s^2}{9}-\frac{10\s^3}{27}\right)
\right]\e_\V^2\nonumber\\
&&+\left[-\frac{14\s\a_0}{3}+\frac{4\s\nu_0}{3}\left(1-\frac{\s}{3}\right) \right] \eta_\V^2 
\nonumber \\
& & +\left[2\a_0\left(-16+\frac{46\s}{3}-\frac{\s^2}9\right)+\nu_0\left(32-\frac{70\s}{3}+\frac{13\s^2}{9}-\frac{4\s^3}{27}\right)\right]\e_\V \eta_\V \nonumber \\
& & +\left[2\a_0\left(2-\frac{\s}{3}\right)+4\nu_0\left(\frac{2\s}{3}-1\right)\right] \xi_\V^2\,,\\
f_{\rm s}&:=& \frac{\sigma [3\a_0(13\s-3)+ \nu_0\s(6+11\s)]}{18(\sigma+1)}\,.
\ea\es
The tensor index \Eq{nt} and its running \Eq{alt} are
\ba
\label{ntf}
n_{\rm t} &=& -2\e_\V-c_{n_{\rm t}}\dpl\,,\label{nt2} \\
\label{alphatf}
\alpha_{\rm t} &=& -4 \e_\V \left( 2\e_\V-\eta_\V \right)
+c_{\a_{\rm t}}\dpl\,,
\ea
where
\ba
c_{n_{\rm t}} &=& 
f_{\rm t}-\left[2\alpha_0 (1-\sigma)
+\nu_0 (\sigma-2) \right] \e_\V-\frac{2\sigma\alpha_0}{3} \eta_\V \,,\\
c_{\a_{\rm t}} &=&
\s f_{\rm t}+\s[(2-\s)\nu_0-2\alpha_0]\e_\V
+\left[16\a_0\left(1-\s\right)-4\nu_0\left(4-\frac{8\s}{3}+\frac{\s^2}{9}\right)\right]\e_\V^2 \nonumber \\
& &-\frac{4\s\a_0}3\eta_\V^2
+\left[2\a_0\left(5\s-4\right)+2\nu_0\left(4-\frac{7\s}{3}\right)\right]\e_\V \eta_\V\,,\\
f_{\rm t}&:=& \frac{2\sigma^2 \alpha_0}{\sigma+1}\,.
\ea
Finally, the tensor-to-scalar ratio (\ref{tts2}) is 
\be\label{str1}
r=16\e_\V+c_r\dpl\,,
\ee
where 
\be\label{str2}
c_r=\frac{8[3\a_0(3+5\s+6\s^2)-\nu_0\s(6+11\s)]}{9(\sigma+1)}\,\e_\V-\frac{16\s\a_0}{3} \eta_\V\,.
\ee

\section{Power spectra and cosmic variance}\label{ksr}

In this section, we cast the power spectra as nonperturbative functions of the wavenumber $k$ 
and a pivot scale $k_0$ (section \ref{ksr1}). The parameter space of the numerical analysis 
is introduced in section \ref{ksr2}, while a theoretical prior on the size of the quantum correction 
is discussed in section \ref{ksr3}. An important question to address is whether a possible LQC 
signal at large scales would be stronger than cosmic variance, which is the dominant effect 
at low multipoles. This issue is considered in section \ref{ksr4}, where a positive answer is 
given for a certain range in the parameter space.

\subsection{Power spectra and pivot scales}\label{ksr1}

Because of eq.~\Eq{dDd}, terms higher than 
the runnings $\alpha_{\rm s}$ and $\alpha_{\rm t}$ can give rise to a nonnegligible contribution 
to the power spectra ${\cal P}_{\rm s} (k)$ and ${\cal P}_{\rm t} (k)$.
Let us expand the scalar  power spectrum to all orders in the perturbation wavenumber 
about a pivot scale $k_0$:
\be
\ln {\cal P}_{\rm s} (k)=\ln {\cal P}_{\rm s} (k_0)+[n_{\rm s}(k_0)-1] x
+\frac{\alpha_{\rm s}(k_0)}{2}x^2
+\sum_{m=3}^{\infty} \frac{\alpha_{\rm s}^{(m)}(k_0)}{m!}x^m\,,
\label{Ps2}
\ee
where $x := \ln (k/k_0)$, and $\alpha_{\rm s}^{(m)}:= \rmd^{m-2} \a_{\rm s}/(\rmd\ln k)^{m-2}$.
When $O(\e_\V)$ and $O(\eta_\V)$ terms are ignored, $c_{n_{\rm s}}\approx f_{\rm s}$ in eq.~(\ref{ns2}), 
while the dominant contribution to the scalar running can be estimated as
\be
\alpha_{\rm s} (k_0)=\frac{\rmd n_{\rm s}}{\rmd \ln k}\biggr|_{k=k_0}
\approx \sigma f_{\rm s} \dpl (k_0)\,.
\ee
Similarly, we can derive the $m$-th order terms $\alpha_{\rm s}^{(m)}$ as
\be\label{aln}
\alpha_{\rm s}^{(m)} (k_0) \approx (-1)^m \s^{m-1}f_{\rm s}\dpl (k_0)\,.
\ee
In this case, the last term in eq.~(\ref{Ps2}) converges to the exponential series,
\be
\sum_{m=3}^{\infty} \frac{\alpha_{\rm s}^{(m)}(k_0)}{m!}x^m
=f_{\rm s}\dpl (k_0) \left[x\left(1-\frac12 \sigma x \right)
+\frac{1}{\sigma} (e^{-\sigma x}-1)\right]\,.
\ee
Thus, the scalar power spectrum (\ref{Ps2}) can be written in the form 
\ba
{\cal P}_{\rm s} (k)&=&{\cal P}_{\rm s} (k_0) \exp \left\{ [n_{\rm s}(k_0)-1]x
+\frac{\alpha_{\rm s}(k_0)}{2}
x^2\right.\nonumber\\
&&\left.+f_{\rm s} \dpl(k_0) \left[x \left(1-\frac12 \sigma x \right)
+\frac{1}{\sigma} (e^{-\sigma x}-1) \right] \right\}\,.
\label{Psfinal}
\ea

We stress that the power spectrum (\ref{Psfinal}) is nonperturbative in the wavenumber $k$, 
while ordinary inflationary analyses take a truncation of (\ref{Ps2}).
The expression (\ref{Psfinal}) is valid for any value of $\sigma$ and of the pivot wavenumber, 
provided the latter lies within the observational range of the experiment. 
Note that $k_0$ is not fixed observationally, except from the fact that we can choose any value 
on the scales relevant to CMB (with the multipoles $\ell$ ranging 
in the region $2<\ell<1000$). 
The CMB multipoles are related to the wavenumber $k$ 
by the approximate relation
\begin{equation}
k\approx 10^{-4}h\,\ell~~{\rm Mpc}^{-1}\,,
\end{equation}
where we take the value $h=0.7$ for the reduced Hubble constant. 
The default pivot value of {\sc cmbfast} \cite{cmbfast}
and {\sc camb} \cite{Antony} codes is $k_0=0.05$~Mpc$^{-1}$ ($\ell_0\sim 730$). 
For the WMAP pivot scale $k_0=0.002$~Mpc$^{-1}$ ($\ell\sim 29$) \cite{Peiris,Komatsu}, 
the maximum value of $x$ relevant to the CMB anisotropies is $x_{\rm max} \sim 3.6$. 
Intermediate values of $k_0$ are also possible, for instance $k_0=0.01$~Mpc$^{-1}$ \cite{Leach}. 
In general, the constraints on the parameter space, and in particular the likelihood contours, 
depend (even strongly) on the choice of the pivot scale \cite{CLM}, and it is interesting 
to compare results with different $k_0$ also in LQC.

The fact that we can resum the whole series is of utmost importance for the consistency of 
the numerical analysis. In standard inflation, higher-order terms do not contribute to the power 
spectrum because they are higher-order in the slow-roll parameters. Then, one can truncate 
eq.~\Eq{Ps2} to the first three terms and ignore the others. Here, on the other hand, all the 
terms \Eq{aln} are linear in $\dpl$ and they contribute equally if the parameter $\s$ is large 
enough, $\sigma\gtrsim 1$. This fact might naively suggest that small values of $\sigma$ 
are preferred for a consistent analysis of a quasi-scale-invariant spectrum \cite{InflObs}. 
In that case, one would have to impose conditions such as 
$|[n_{\rm s}(k_0)-1]x| \gg |[\alpha_{\rm s}(k_0)/2]x^2|$, 
which depend on the pivot scale $k_0$. 

For $\s>1$, however, different choices of $k_0$ would result in different convergence 
properties of the Taylor expansion of ${\cal P}_{\rm s}$. The point is that $\dpl(k)$ 
changes fast for $\s>1$ and the running of the spectral index can be sizable; dropping 
higher-order terms would eventually lead to inconsistent results. 
On the other hand, eq.~\Eq{Psfinal} does not suffer from any of the above
limitations and problems, and it will be the basis of our analysis, where $n_{\rm s}(k_0)$
 and $\alpha_{\rm s}(k_0)$ are given by eqs.~(\ref{ns2}) and (\ref{alphas}). 
 The last term in eq.~\Eq{Psfinal}, usually negative, tends to compensate the large 
 positive running, thus providing a natural scale-invariance mechanism without putting 
 any numerical priors.

Assuming that $c_{n_{\rm t}}\approx f_{\rm t}$, same considerations hold for 
the tensor spectrum, which can be written as
\ba
{\cal P}_{\rm t} (k)&=&{\cal P}_{\rm t} (k_0) \exp \left\{ n_{\rm t} (k_0)\,x
+\frac{\alpha_{\rm t}(k_0)}{2} x^2\right.\nonumber\\
&&\qquad\qquad\qquad\left.+f_{\rm t} \dpl(k_0) \left[x \left(1-\frac12 \sigma x \right)
+\frac{1}{\sigma} (e^{-\sigma x}-1) \right] \right\}\,,
\label{Ptfinal}
\ea
where $n_{\rm t} (k_0)$ and $\alpha_{\rm t} (k_0)$ are given by eqs.~(\ref{ntf})
and (\ref{alphatf}), respectively. Finally, the tensor-to-scalar ratio is given by 
eqs.~\Eq{str1} and \Eq{str2}, with the slow-roll parameters 
evaluated at the pivot scale $k=k_0$.

\subsection{Parameter space}\label{ksr2}

The CMB likelihood analysis can be carried out by using 
eqs.~(\ref{Psfinal}), (\ref{Ptfinal}), and (\ref{str1}).
Let us take the power-law potential \cite{Linde}
\be
V(\vp)=V_0 \vp^n\,.
\label{powerpo}
\ee
In this case, it follows that ($k_0$ dependence implicit) 
\be
\e_\V=\frac{n^2}{2\kappa^2 \vp^2}\,,
\qquad
\eta_\V=\frac{2(n-1)}{n}\e_\V\,,
\qquad
\xi_\V^2=\frac{4(n-1)(n-2)}{n^2}\e_\V^2\,.
\ee
This allows us to reduce the slow-roll parameters to 
one (i.e., $\epsilon_\V$).

For the exponential potential \cite{Mata}
\be
V(\vp)=V_0 e^{-\kappa \lambda \vp}\,,
\label{exppo}
\ee
the relation between the slow-roll parameters is given by 
\be
\e_\V=\frac{\lambda^2}{2}\,,
\qquad
\eta_\V=2\e_\V\,,
\qquad
\xi_\V^2=4\e_\V^2\,,
\ee
which are again written in terms of the single parameter
$\e_\V$.

Between the model parameters $\nu_0$ and $\alpha_0$ we can 
also impose the following relation \cite{InflObs}, valid for $\s\neq 3$:
\be
\nu_0=\frac{3(\sigma-6)}{(\sigma+6)(\sigma-3)}\alpha_0\,.
\label{nure}
\ee
Introducing the variable 
\be\label{deltaa}
\delta (k_0) := \alpha_0 \dpl(k_0)\,,
\ee
we can write $f_{\rm s} \dpl (k_0)$ and $f_{\rm t} \dpl (k_0)$ in the form
\be
f_{\rm s} \dpl (k_0)=\frac{\sigma(8\sigma^3-8\sigma^2-93\sigma+18)}
{2(\sigma-3)(\sigma+1)(\sigma+6)}\delta (k_0)\,,\qquad
f_{\rm t} \dpl (k_0)=\frac{2\sigma^2}{\sigma+1}\delta (k_0)\,.
\ee
For $\s=3$ one has $\a_0=0$ identically, in which case 
eq.~\Eq{deltaa} is replaced by 
$\delta (k_0) := \nu_0 \dpl(k_0)$.

To summarize, using the relation (\ref{nure}), all the other observables 
can be written in terms of $\delta(k_0)$ and $\e_\V(k_0)$. 
Hence, for given $\sigma$ and $k_0$, 
one can perform the CMB likelihood analysis by varying the 
two parameters $\delta (k_0)$ and $\e_\V(k_0)$.

\subsection{Theoretical upper bound on the quantum correction}\label{ksr3}

For the validity of the linear expansion of the correction functions
\Eq{an} and \Eq{cn}\footnote{Using the relation (\ref{nure}), one sees 
that $\nu_0$ is of the same order as $\a_0$, so a bound on $\delta$ is sufficient.} 
and all the perturbation formulas where the $O(\dpl)$ truncation has been 
systematically implemented, we require that $\delta(k)=\alpha_0 \dpl(k)<1$ 
for all wavenumbers relevant to the CMB anisotropies.
Since $\dpl \propto a^{-\sigma}$, the quantity $\delta(k)$ 
appearing in inflationary observables is approximately given by 
\be
\delta(k)=\delta(k_0) \left(\frac{k_0}{k}\right)^{\sigma}\,,
\label{deltak}
\ee
where we have used $k=\cH$ at Hubble exit with $\cH/a \approx {\rm const}$. 
As $k\propto\ell$, the same expression can be written in terms of 
the multipoles $\ell$. Since $\sigma>0$, one has $\delta (k)>\delta (k_0)$ 
for $k<k_0$ and $\delta (k)<\delta (k_0)$ for $k>k_0$. 
This means that the larger the pivot scale $k_0$, 
the smaller the upper bound on $\delta (k_0)$.

Let us consider two pivot scales: 
(i) $k_0=0.002$~Mpc$^{-1}$ (multipole $\ell_0\sim 29$)
and 
(ii) $k_0=0.05$~Mpc$^{-1}$ (multipole $\ell_0\sim 730$). 
Since the largest scale in
CMB corresponds to the quadrupole $\ell=2$, the condition $\delta(k)<1$ at $\ell=2$ 
gives the following bounds $\delta_{\rm max}$ on the values of $\delta(k_0)$ with two pivot scales:
\ba
& &{\rm (i)}~~\delta_{\rm max}=
14.5^{-\sigma} \qquad ({\rm for}~~k_0=0.002~{\rm Mpc}^{-1})\,,\\
& &{\rm (ii)}~~\delta_{\rm max}=
365^{-\sigma} \qquad ({\rm for}~~k_0=0.05~{\rm Mpc}^{-1})\,.
\label{delmax2}
\ea
Values of $\delta_{\rm max}$ for some choices of $\s$ are reported 
in table  \ref{tab1}. The suppression of $\delta_{\rm max}$ for larger $k_0$ 
and $\sigma$ can be also seen in the power spectra (\ref{Psfinal}) and (\ref{Ptfinal}). 
The term $e^{-\sigma x}=(k_0/k)^\s$ can be very large for large $k_0$: for instance, 
if $\sigma=6$ and $k_0=0.05$~Mpc$^{-1}$, one has $e^{-\sigma x} \sim 10^{15}$ 
at $\ell=2$. Then we require that 
$\delta (k_0)$ is suppressed as $\delta(k_0) \lesssim 10^{-16}$.

\TABLE{
\begin{tabular}{|c||c|c|c|c|c|c|}
\hline
$\sigma$ & 0.5 & 1 & 1.5 & 2 & 3 & 6  \\
\hline\hline
\multicolumn{7}{|c|}{$k_0=0.002~{\rm Mpc}^{-1}$}\\\hline
$\delta_{\rm max}$ & 0.26 & $6.9 \times 10^{-2}$ & $1.8 \times 10^{-2}$ 
& $4.7 \times 10^{-3}$ & $3.2 \times 10^{-4}$ & $1.0 \times 10^{-7}$ \\
$\delta$ & 0.27 & $3.5 \times 10^{-2}$ & $1.7 \times 10^{-3}$ & $6.8 \times 10^{-5}$ 
& $4.3 \times 10^{-7}$ & -- \\
\hline\hline
\multicolumn{7}{|c|}{$k_0=0.05~{\rm Mpc}^{-1}$}\\\hline
$\delta_{\rm max}$ & $5.2 \times 10^{-2}$ & $2.7 \times 10^{-3}$ & 
$1.4 \times 10^{-4}$ & $7.5 \times 10^{-6}$ & $2.1 \times 10^{-8}$ &
$4.3 \times 10^{-16}$ \\
$\delta$ & $6.7 \times 10^{-2}$ & $9.0 \times 10^{-4}$ & $1.3 \times 10^{-5}$ 
& $1.2 \times 10^{-7}$ 
& $ 2.7 \times 10^{-11}$  & -- \\
\hline
\end{tabular}
\caption{\label{tab1} 
Theoretical priors on the upper bound of $\delta$ ($=\delta_{\rm max}$) 
and 95\% CL upper limits of $\delta$ constrained by observations for 
the potential $V(\varphi)=V_0 \varphi^2$
with different values of $\sigma$ and for two pivot scales. 
The likelihood analysis has not been performed for $\s=6$ 
since the signal is below the cosmic variance threshold already when $\s=2$. For $\s=3$, the parameter $\delta=\nu_0\dpl$ has been used.}
}

\subsection{Cosmic variance}\label{ksr4}

At large scales, the failure of the ergodic theorem for the CMB multipole 
spectrum manifests itself in the phenomenon of cosmic variance,
an intrinsic uncertainty on observations due to the small samples at low multipoles. 
For a power spectrum ${\cal P}(\ell)$, cosmic variance is given by \cite{variance}
\be
{\rm Var}_{{\cal P}}(\ell)=\frac{2}{2\ell+1}\,{\cal P}^2(\ell)\,.
\ee
A natural question, which is often overlooked in the literature of exotic cosmologies, 
is how effects coming from new physics compete with cosmic variance. 
In our particular case, we would like to find which values of $\s$ give rise to 
a theoretical upper bound $\delta_{\rm max}$ of inverse-volume LQC 
quantum corrections larger than the error bars due to 
cosmic variance \emph{with respect to the classical spectrum}.

Consider the scalar spectrum ${\cal P}_{\rm s} (\ell)$, eq.~\Eq{Psfinal} 
with $k/k_0$ replaced by $\ell/\ell_0$. It is determined up to the 
normalization ${\cal P}_{\rm s} (\ell_0)$, so that the region in the 
$(\ell,{\cal P}_{\rm s} (\ell)/{\cal P}_{\rm s} (\ell_0))$ plane 
affected by cosmic variance is roughly delimited by the two curves
\be
\frac{{\cal P}_{\rm s} (\ell)\pm \sqrt{{\rm Var}_{{\cal P}_{\rm s}}(\ell)}}
{{\cal P}_{\rm s} (\ell_0)}\Big|_{\dpl=0}=\left(1\pm\sqrt{\frac{2}
{2\ell+1}}\right)\frac{{\cal P}_{\rm s} (\ell)}
{{\cal P}_{\rm s} (\ell_0)}\Big|_{\dpl=0}\,,\label{cosv}
\ee
where we take the classical spectrum as reference.

The power spectrum \Eq{Psfinal}, together with the cosmic variance effect \Eq{cosv},
 is shown in figure \ref{spectrum1} for $n=2$ and the pivot scale $\ell_0=29$. Ignoring the solid lines for the moment, the dashed curves correspond to $\delta=\delta_{\rm max}$. The exponential term $e^{-\sigma x}=(k_0/k)^\sigma$ in eq.~(\ref{Psfinal}) gives rise to an enhancement of the power spectra on large scales, 
as we see in the figures.\footnote{This feature is similar to results for tensor modes \cite{InvVolTens}.} For $\sigma\gtrsim 3$, the growth 
of this term is so significant that $\delta(\ell)$ must be very much 
smaller than 1 for most of the scales observed in the CMB, 
in order to satisfy the bound $\delta (\ell=2) < 1$. LQC inverse-volume corrections are well within the cosmic variance 
 region for $\s>2$. However, already at $\s=2$ quantum corrections strongly 
 affect multipoles $\ell\leq 6$. For $\s\sim 1$, the spectrum is appreciably 
 modified also at multipoles $\ell\gtrsim 500$.
Changing the pivot scale to $\ell_0=730$, one sees that the quantum effect is 
generally greater than cosmic variance at sufficiently low multipoles 
(figure \ref{spectrum2}). The plots for $n=4$ are very similar and we shall omit them.

\FIGURE{
\includegraphics[width=8.0cm]{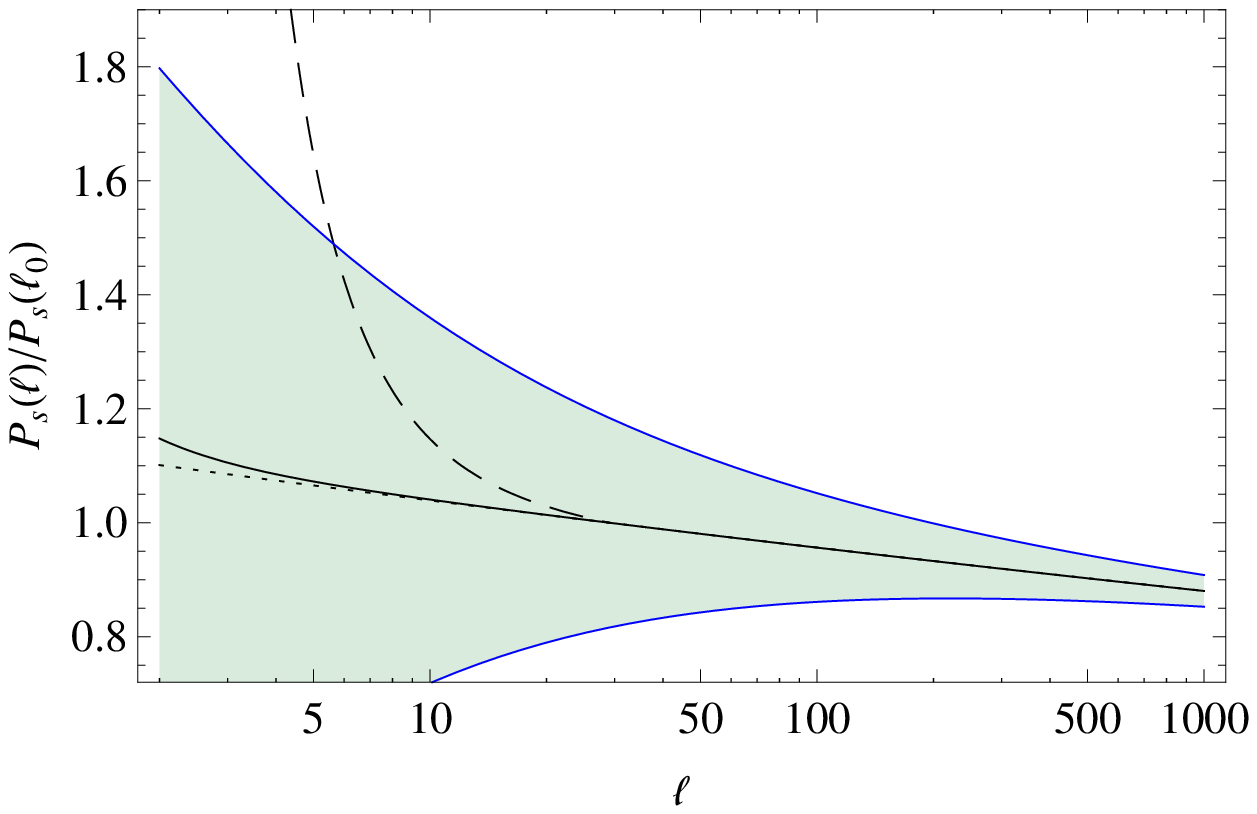}
\includegraphics[width=8.0cm]{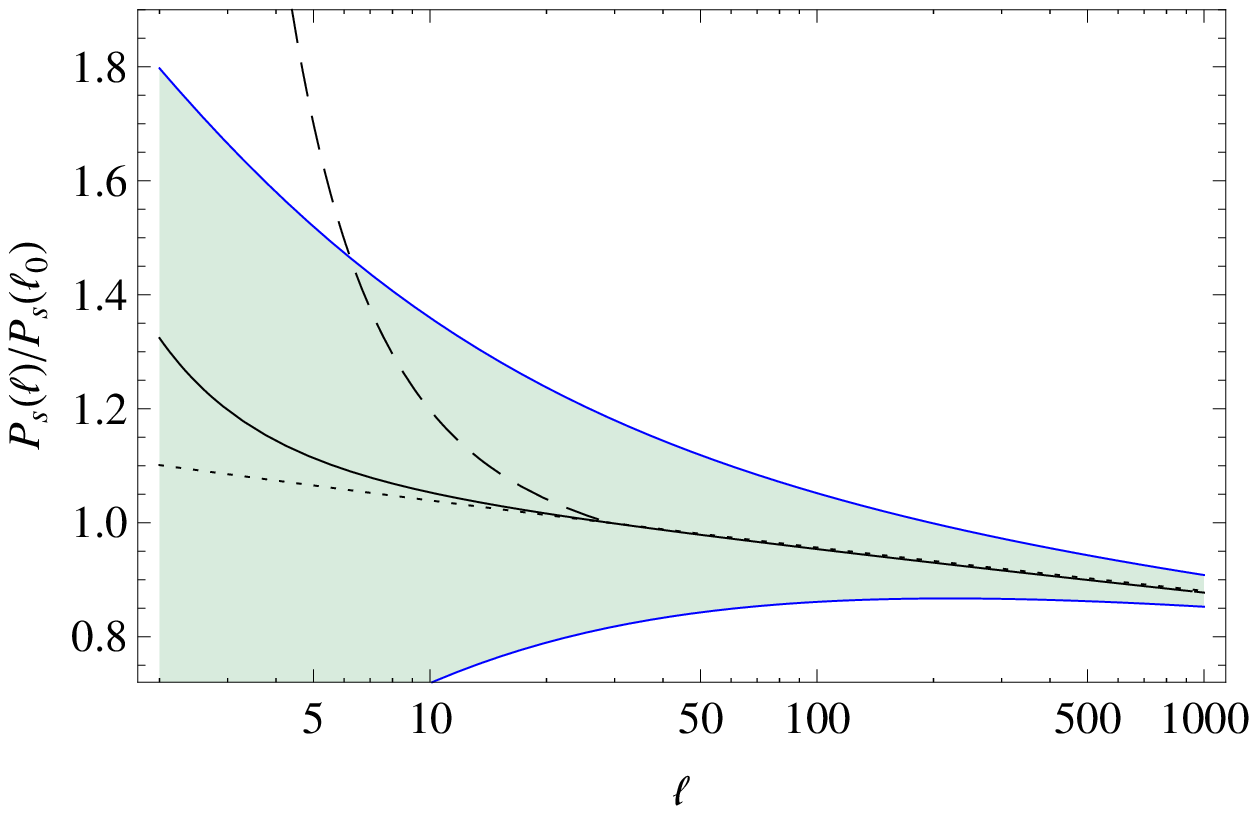}
\includegraphics[width=8.0cm]{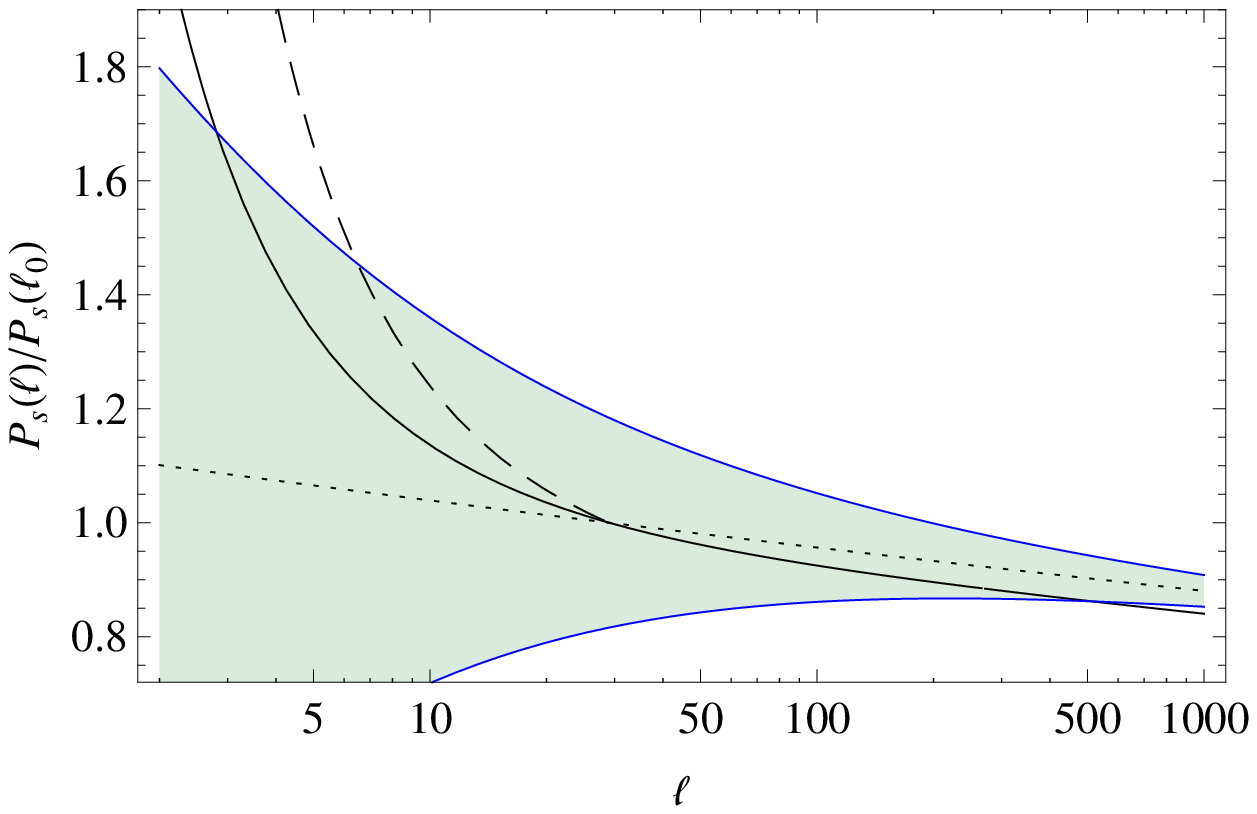}
\caption{\label{spectrum1} Primordial scalar power spectrum ${\cal P}_{\rm s}(\ell)$
for the case $n=2$, with $\epsilon_{\V}(k_0)=0.009$ and the pivot wavenumber 
$k_0=0.002$\,Mpc$^{-1}$, corresponding to $\ell_0= 29$. 
The values of $\s$ are $\s=2$ (top panel) $\s=1.5$ (center panel), 
and $\s=1$ (bottom panel), 
while we choose three different values of $\delta(\ell_0)$, 
as given in table  \ref{tab1}: 0 (classical case, dotted lines), 
the observational upper bound from the numerical analysis (solid lines), 
and $\delta_{\rm max}$ (a-priori upper bound, dashed lines). Shaded regions are affected by cosmic variance.}}

\FIGURE{
\includegraphics[width=8.0cm]{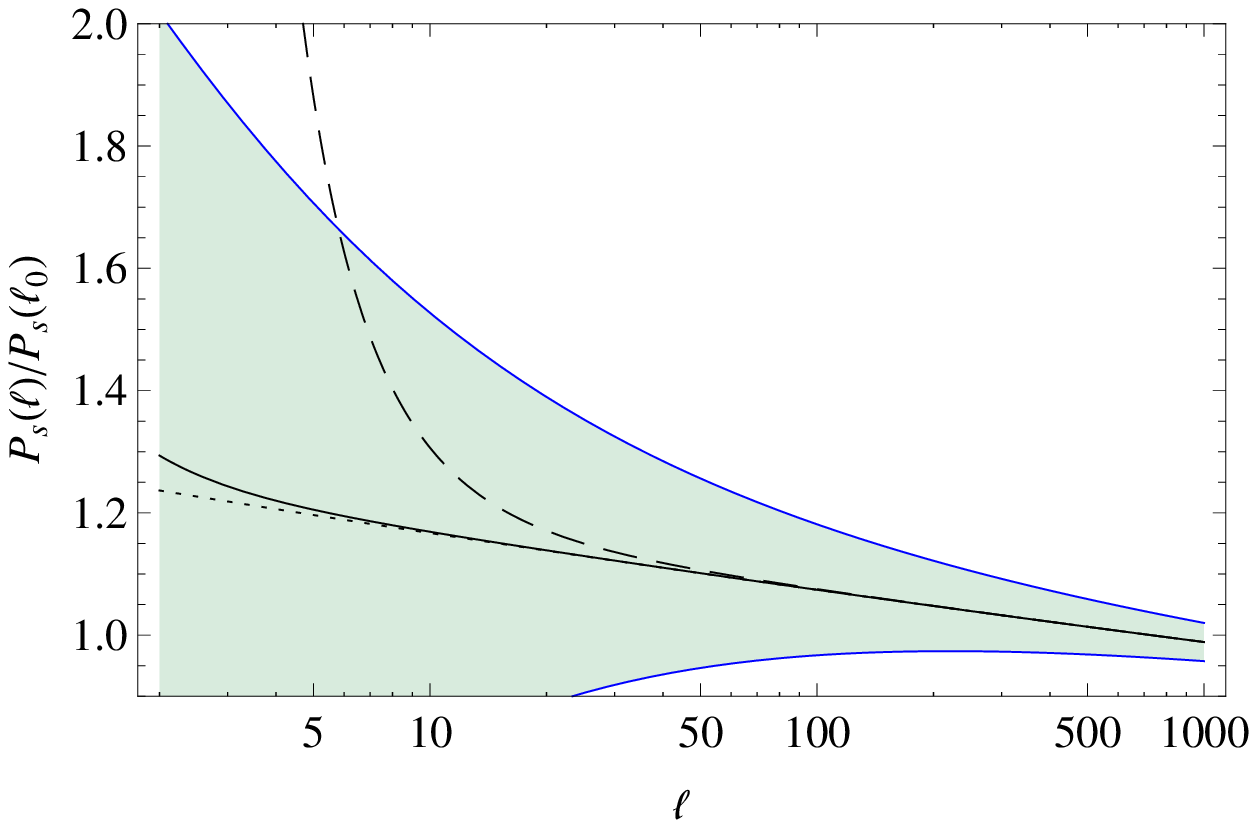}
\includegraphics[width=8.0cm]{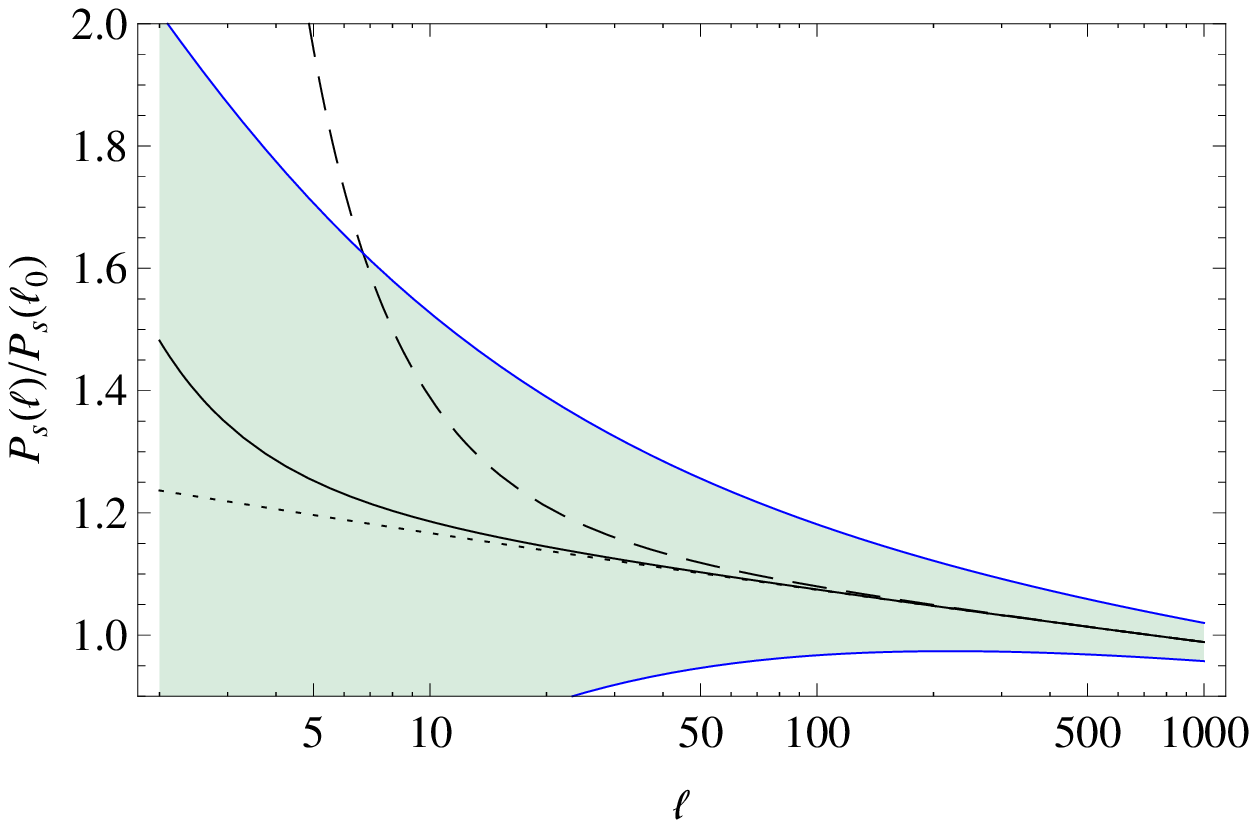}
\includegraphics[width=8.0cm]{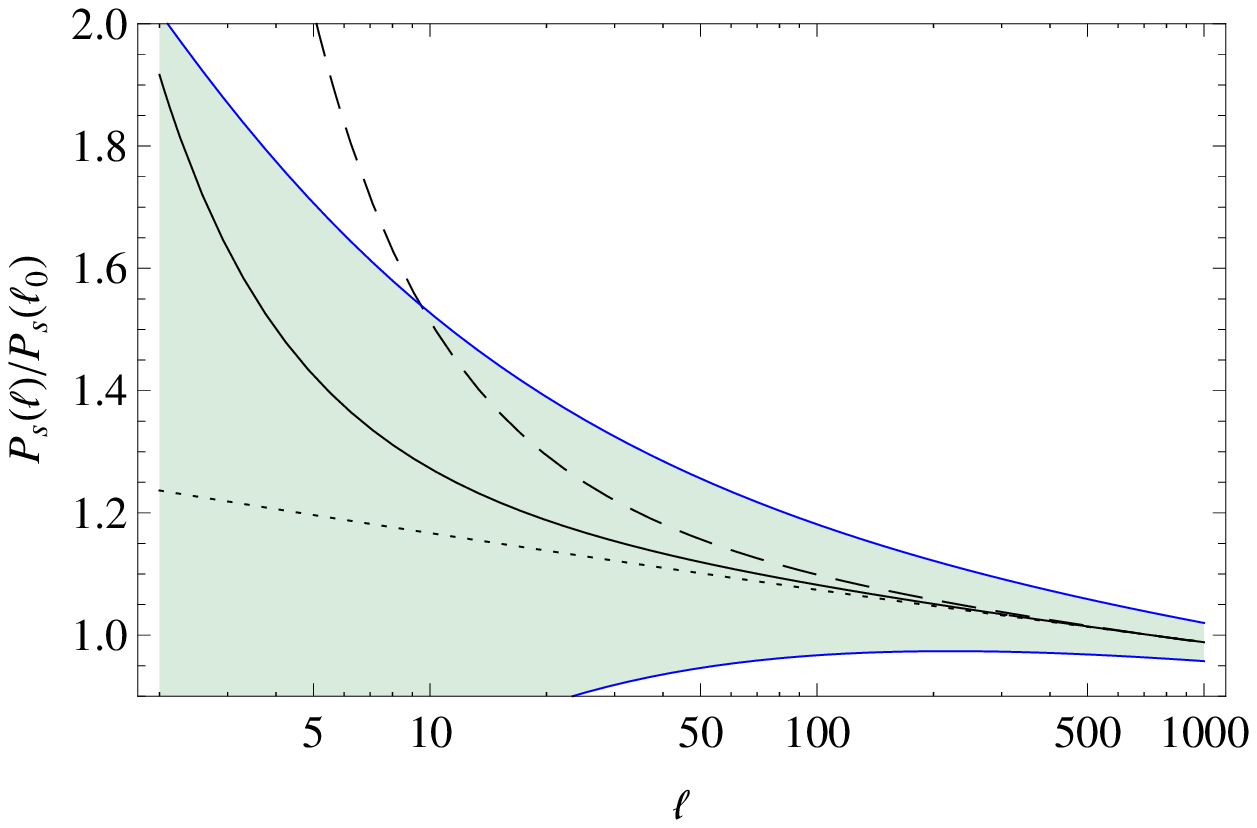}
\caption{\label{spectrum2} Primordial scalar power spectrum ${\cal P}_{\rm s}(\ell)$
for the case $n=2$, with $\epsilon_{\V}(k_0)=0.009$ and 
the pivot wavenumber $k_0=0.05$\,Mpc$^{-1}$, corresponding 
to $\ell_0=730$. The values of $\s$ are $\s=2$ (top panel), $\s=1.5$ (center panel), 
and $\s=1$ (bottom panel), while we choose three different values of $\delta(\ell_0)$, 
as given in table  \ref{tab1}: 0 (classical case, dotted lines), the observational upper 
bound from the numerical analysis (solid lines), and $\delta_{\rm max}$
(a-priori upper bound, dashed lines). Shaded regions are affected by cosmic variance.
}}
\section{Likelihood analysis}
\label{likelihoodsec}

We carry out the CMB likelihood analysis for the power-law 
potential (\ref{powerpo}) as well as the exponential potential (\ref{exppo}).
We run the Cosmological Monte Carlo (CosmoMC) code \cite{Antony} with 
the data of WMAP 7yr \cite{Komatsu} combined with large-scale 
structure (LSS) \cite{Reid} (including BAO), HST \cite{HST}, 
Supernovae type Ia (SN Ia) \cite{SNIa}, and Big
Bang Nucleosynthesis (BBN) \cite{BBN}, by assuming a $\Lambda$CDM
model. 

In the Monte Carlo routine we vary two inflationary parameters 
$\delta (k_0)$ and $\e_\V (k_0)$ as well as other cosmological 
parameters. Note that $\delta (k_0)$ and $\e_\V (k_0)$ are 
constrained at the chosen pivot scale $k_0$. 
While the bound on $\delta$ depends on $k_0$ 
(and it tends to be smaller for larger $k_0$),
that on $(k_0)^\s\delta(k_0)$ does not 
[see eq.~(\ref{deltak})].

Under the conditions $\epsilon_\V \ll 1$ and $\delta \ll 1$,
the slow-roll parameter $\epsilon_\V $ is 
approximately given by 
$\epsilon_\V \approx (\kappa^2/2)(\vp'/\cH)^2$.
Then the number of e-foldings during inflation can be estimated as
\be
N := \int_{\tau}^{\tau_\f} \rmd\tilde{\tau}\,\cH \approx \kappa 
\int_{\varphi_\f}^{\varphi} \rmd\tilde{\varphi} 
\frac{1}{\sqrt{2\e_\V (\tilde{\varphi})}}\,,
\ee
where $\varphi_\f$ is the field value at the end of inflation 
determined by the condition $\epsilon_\V \approx O(1)$. 
For the power-law potential (\ref{powerpo}) one has 
$\varphi_\f \approx n/\sqrt{2\kappa^2}$ and 
$N \approx n/(4\epsilon_\V)-n/4$, which gives 
\be
\label{then}
\e_\V \approx \frac{n}{4N+n}\,,\qquad 45<N<65\,.
\ee
The typical values of $N$ for the perturbations relevant to 
the CMB anisotropies are actually around $50<N<60$, but we have
taken the wider range above.
The comparison of this estimate with the experimental 
range of $\e_\V$ will determine the acceptance or 
exclusion of an inflationary model 
for a given $n$.

For the exponential potential (\ref{exppo}) the slow-roll parameter
$\epsilon_\V$ is constant, which means that inflation does not end unless 
the shape of the potential changes after some epoch.
In this case, we do not have constraints on 
$\epsilon_\V$ coming from the information of 
the number of e-foldings in the observational range.

\subsection{Quadratic potential}

Let us study observational constraints in the case of the 
quadratic potential $V(\vp)=V_0 \vp^2$.

\subsubsection{$k_0=0.002$ Mpc$^{-1}$}
%

\FIGURE[t]{
\includegraphics[width=7.0cm]{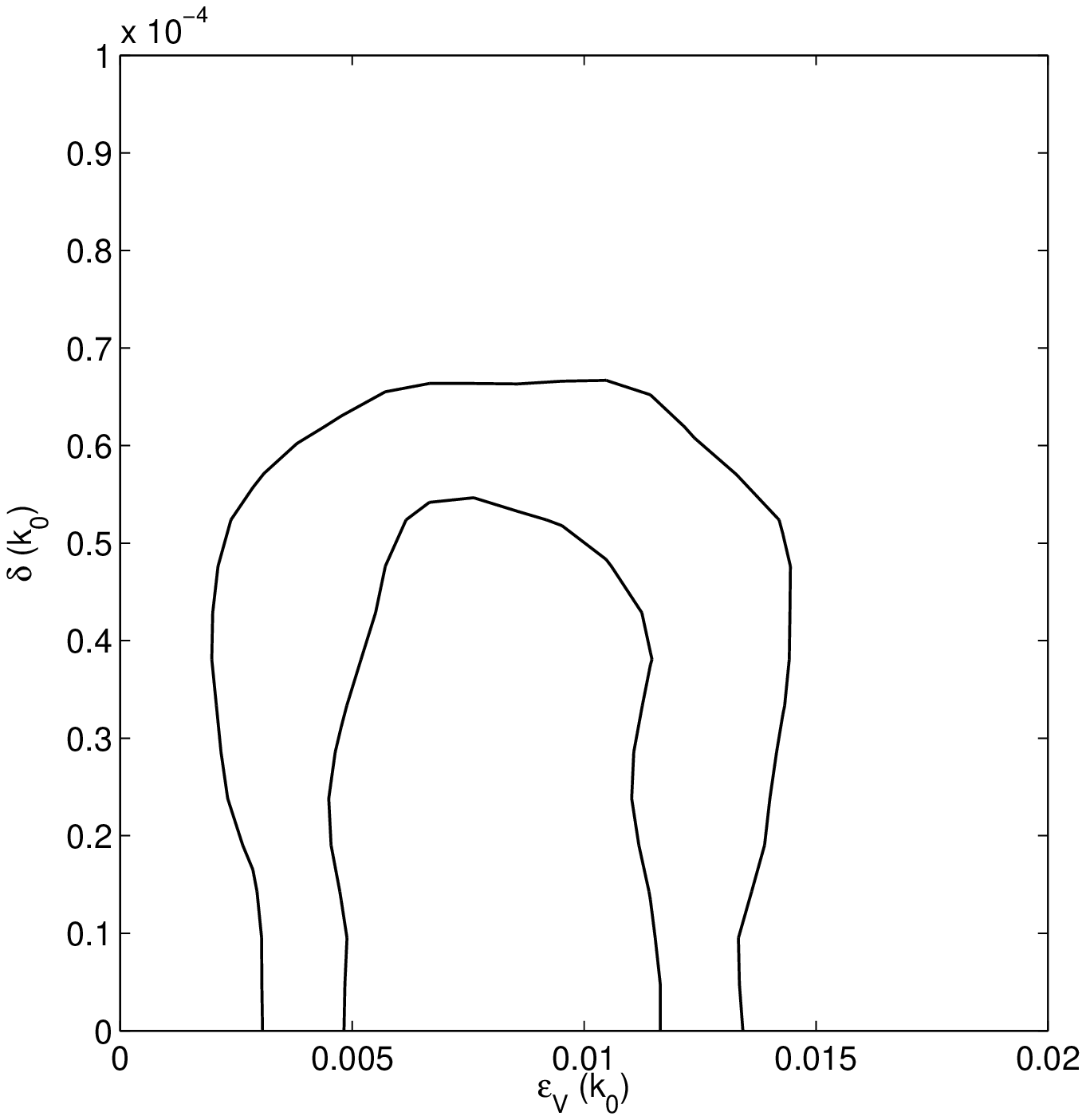}
\includegraphics[width=7.0cm]{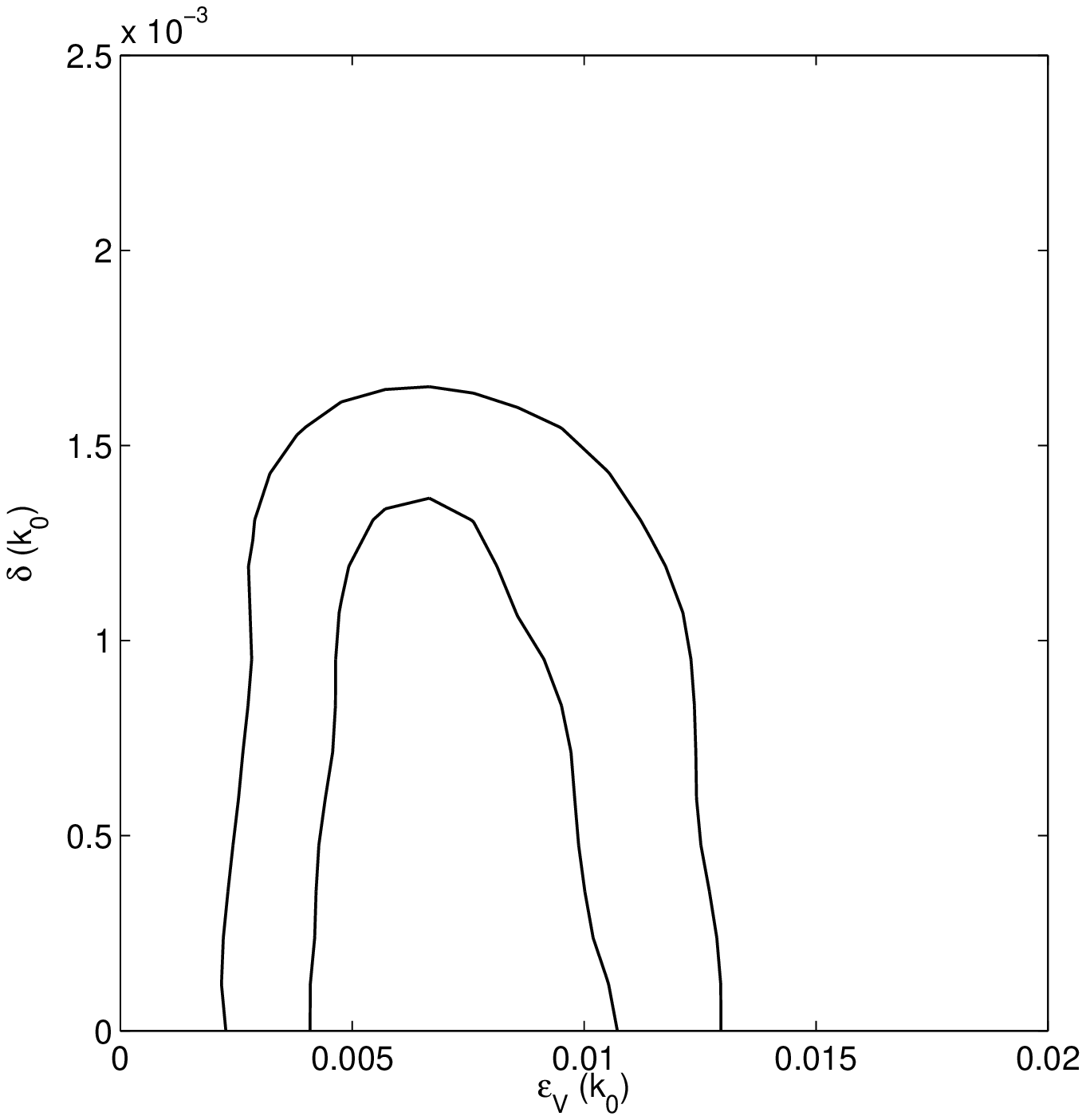}
\includegraphics[width=7.0cm]{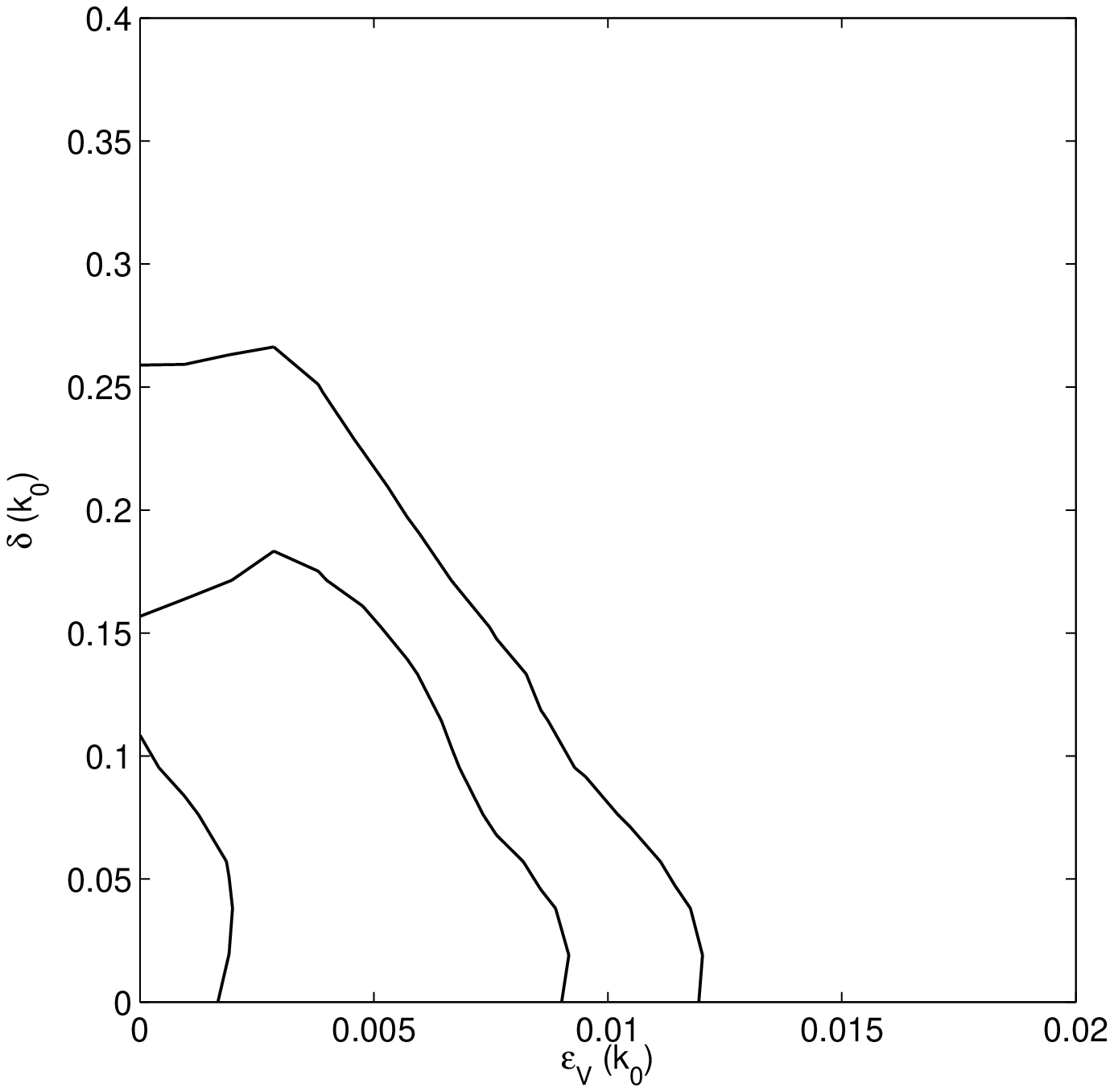}
\caption{2-dimensional marginalized distribution for the quantum-gravity 
parameter $\delta(k_0)$ and the slow-roll parameter $\e_\V (k_0)$ 
with the pivot $k_0=0.002$ Mpc$^{-1}$ for $n=2$,
constrained by the joint data analysis of WMAP 7yr, 
LSS (including BAO), HST, SN Ia, and BBN. 
The values of $\s$ are $\s=2$ (top left panel), 
$\s=1.5$ (top right panel), and $\s=0.5$ (bottom panel). 
The internal and external lines correspond to the 68\% and 95\% 
confidence levels, respectively.\label{en2}}}


We first take the pivot wavenumber $k_0=0.002$ Mpc$^{-1}$ 
($\ell_0\approx 29$) used by the WMAP team \cite{Komatsu}. 
In figure \ref{en2}, the 2D posterior distributions of the parameters 
$\delta (k_0)$ and $\epsilon_\V (k_0)$ are plotted
for $n=2$ and $\sigma=2,1.5,0.5$. 
We have also run the code for other values of $\s$ such 
as $1$ and $3$. The observational upper bounds on $\delta$ 
are given in table \ref{tab1} for several different values of $\s$.

For $\sigma \lesssim 1$, the exponential factor $e^{-\sigma x}$ does not 
change rapidly with smaller values of $f_{\rm s,t}$, so that the
LQC effect on the power spectra would not be very significant 
even if $\delta(k_0)$ was as large as $\e_\V (k_0)$. 
As we see in figure \ref{spectrum1} (solid curve), if $\sigma=0.5$ the LQC correction is constrained to be 
$\delta (k_0)<0.27$ (95\% CL), 
which exceeds the theoretical prior $\delta_{\rm max}=0.26$.
Since $\delta (k_0)$ is as large as 1 in such cases, the validity of 
the approximation $\delta (k_0)<\e_\V (k_0)$ to derive the power
spectra is no longer reliable for $\sigma \lesssim 0.5$.

Looking at table \ref{tab1}, when $\s=1$, the observational upper bound 
on $\delta(k_0)$ becomes of the same order as $\delta_{\rm max}$.
For $\s \lesssim 1.5$ the effect of the LQC correction to the power 
spectrum becomes important on large scales relative to 
cosmic variance. For smaller $\s$ the 
observational upper bound on 
$\delta(k_0)=\alpha_0 \delta_{\rm Pl} (k_0)$ 
tends to be larger.
When $\s=1.5$ the LQC correction is constrained to be
$\delta (k_0)<1.7 \times 10^{-3}$ (95\% CL), see figure \ref{en2}. 
This is smaller than the theoretical prior 
$\delta_{\rm max}=1.8 \times 10^{-2}$ by one order of magnitude.

The effect of cosmic variance is 
significant for $\s \gtrsim 1.5$. 
When $\s=3$, the LQC correction is constrained to be 
$\delta (k_0)=\nu_0 \delta_{\rm Pl} (k_0)<4.3 \times 10^{-7}$ 
(95\% CL).
With respect to the prior $\delta_{\rm max}= 3.3 \times 10^{-4}$, 
the observational bound is smaller by three orders of magnitude.
For $\sigma \gtrsim 3$ the power spectra grow very sharply 
for low multipoles, so that the upper bounds on $\delta(k_0)$
become smaller. Numerically it is difficult to deal with 
such rapidly changing power spectra. 

For the sake of completeness, we should notice that the bounds plotted in Figs.~\ref{spectrum1} and \ref{spectrum2} include input from several datasets, but the cosmic variance belt comes only from the CMB. Therefore, the medium-scale part of the cosmic-variance plots might not give the full picture of the statistical limitations in this range, but we do not expect appreciable modifications from large-scale structure observations.

For $n=2$, the theoretically constrained region \Eq{then} corresponds to 
$0.008<\epsilon_\V<0.011$. 
As we see in figure \ref{en2}, for $\sigma \gtrsim 0.5$, 
the probability distributions of $\epsilon_\V$ are 
consistent with this range even in the presence 
of the LQC corrections. 
Hence, for the pivot wavenumber $k_0=0.002~{\rm Mpc}^{-1}$,
the quadratic potential is compatible with observations 
as in standard cosmology.

\subsubsection{$k_0=0.05$ Mpc$^{-1}$}
%

\FIGURE{
\includegraphics[width=7.0cm]{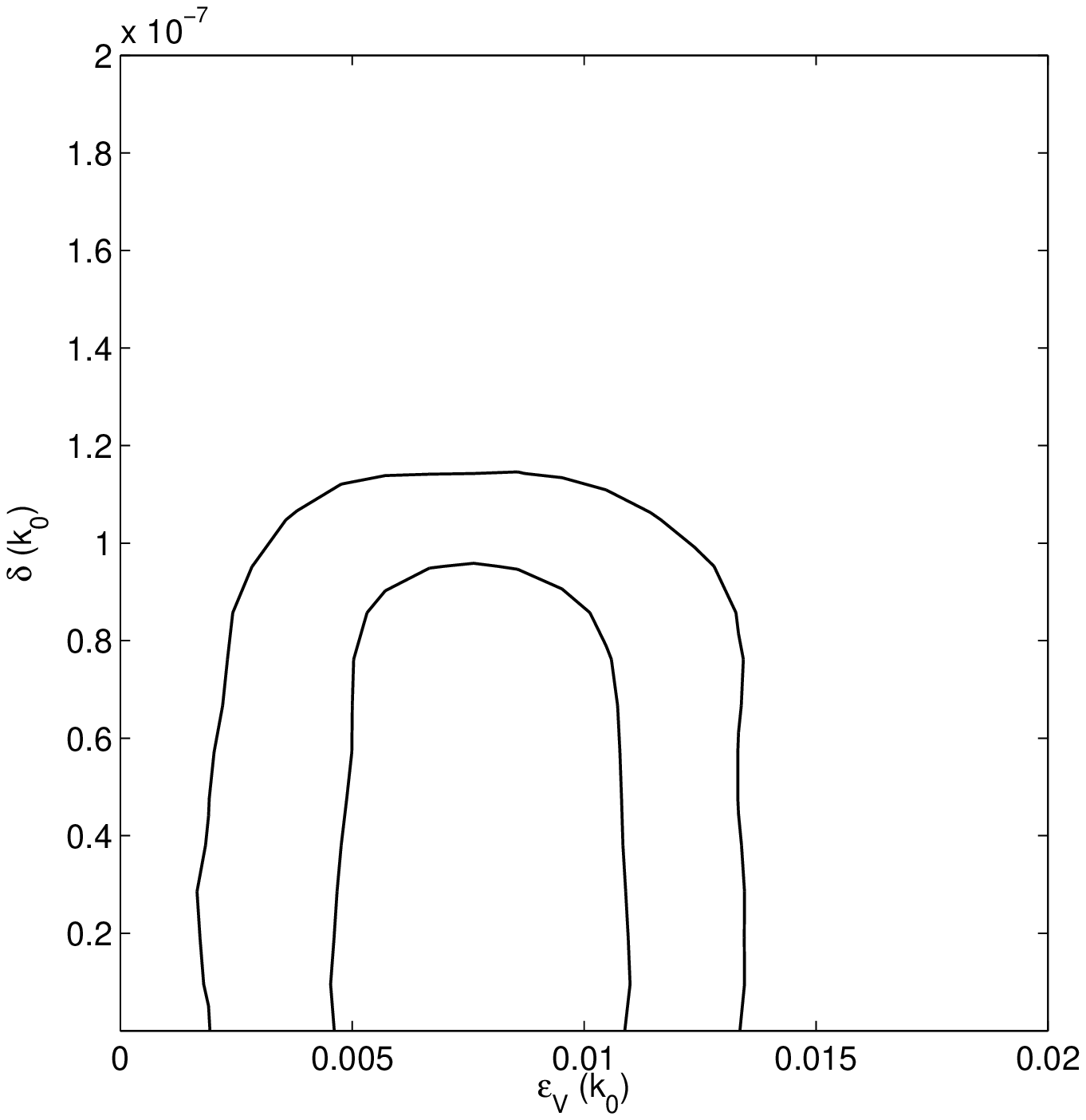}
\includegraphics[width=7.0cm]{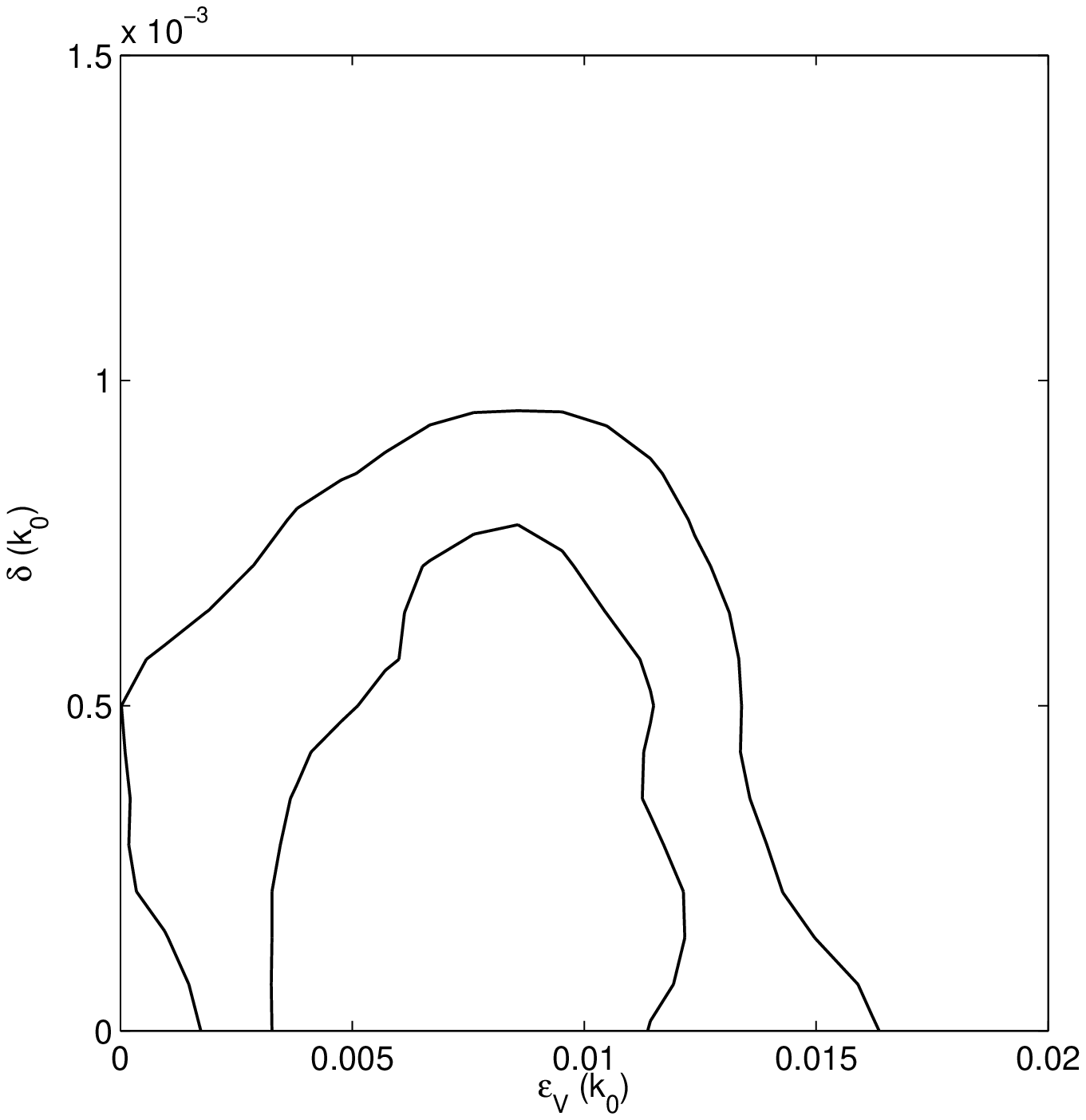}
\caption{2-dimensional marginalized distribution for $n=2$ 
with the pivot $k_0=0.05$ Mpc$^{-1}$.
The values of $\s$ are $\s=2$ (left panel) and 
$\s=1$ (right panel).\label{en2d}}}


We proceed to the case of the pivot wavenumber $k_0=0.05~{\rm Mpc}^{-1}$
($\ell_0 \approx 730$). From eq.~(\ref{delmax2}), the theoretical priors on 
$\delta_{\rm max}$ for given $\s$ are smaller than those corresponding 
to $k_0=0.002~{\rm Mpc}^{-1}$.
In figure \ref{en2d} we plot the 2-dimensional posterior distribution 
of $\delta (k_0)$ and $\epsilon_\V (k_0)$ for 
$n=2$ and $\s=2, 1$. 
When $\s=2$, the observational upper limit is found to be 
$\delta (k_0)<1.2 \times 10^{-7}$ (95\% CL), which 
is two orders of magnitude smaller than the bound 
$\delta (k_0)< 6.8 \times 10^{-5}$
obtained for the pivot $k_0=0.002~{\rm Mpc}^{-1}$.
This comes from the fact that the choice of larger $k_0$ leads to 
more enhancement of power on large scales.

When $\s=1$, we find the constraint
$\delta (k_0)<9.0 \times 10^{-4}$ (95\% CL), which 
is about 1/3 of the theoretical prior 
$\delta_{\rm max}= 2.7 \times 10^{-3}$.
From table  \ref{tab1}, we see that the observational limit of $\delta$ 
for $\s=0.5$ exceeds  $\delta_{\rm max}$.
Hence, our combined slow-roll/$\dpl$ truncation 
is no longer trustable for $\s \lesssim 0.5$, as it happens
for $k_0=0.002~{\rm Mpc}^{-1}$.

{}From figure \ref{en2d} we find that the theoretically allowed 
range of $\epsilon_\V$ ($0.008<\epsilon_\V<0.011$) is
consistent with its observational constraints.
The different choice of $k_0$ affects 
the upper bounds on $\delta (k_0)$, but the basic property 
of the LQC effect on the power spectra is similar.
The quadratic inflaton potential can be consistent with the combined
observational constraints even in the presence of 
the LQC corrections, independent of the values of $k_0$ 
relevant to the CMB anisotropies.

\subsection{Quartic potential}

\FIGURE{
\includegraphics[width=7.0cm]{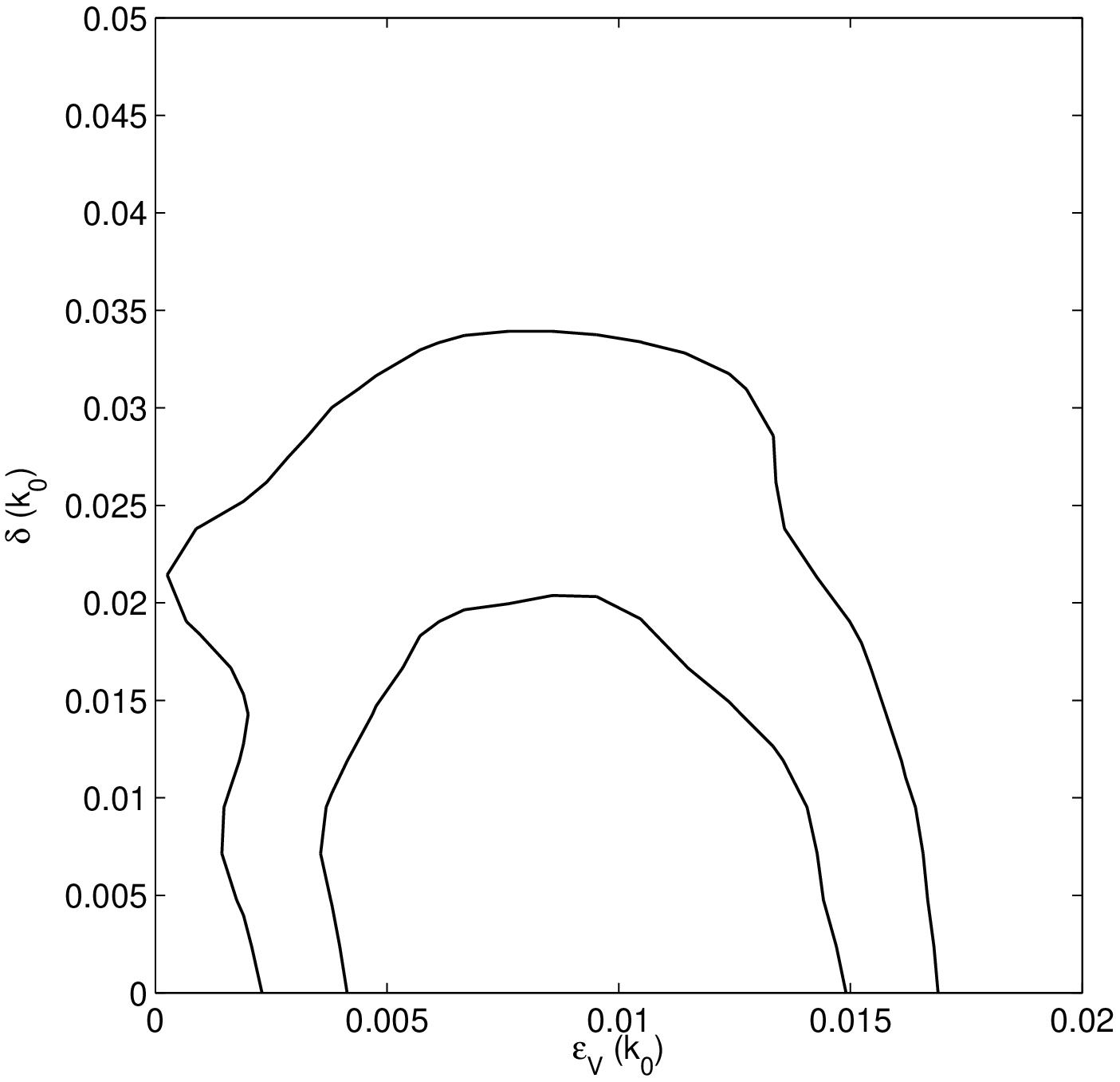}
\includegraphics[width=7.0cm]{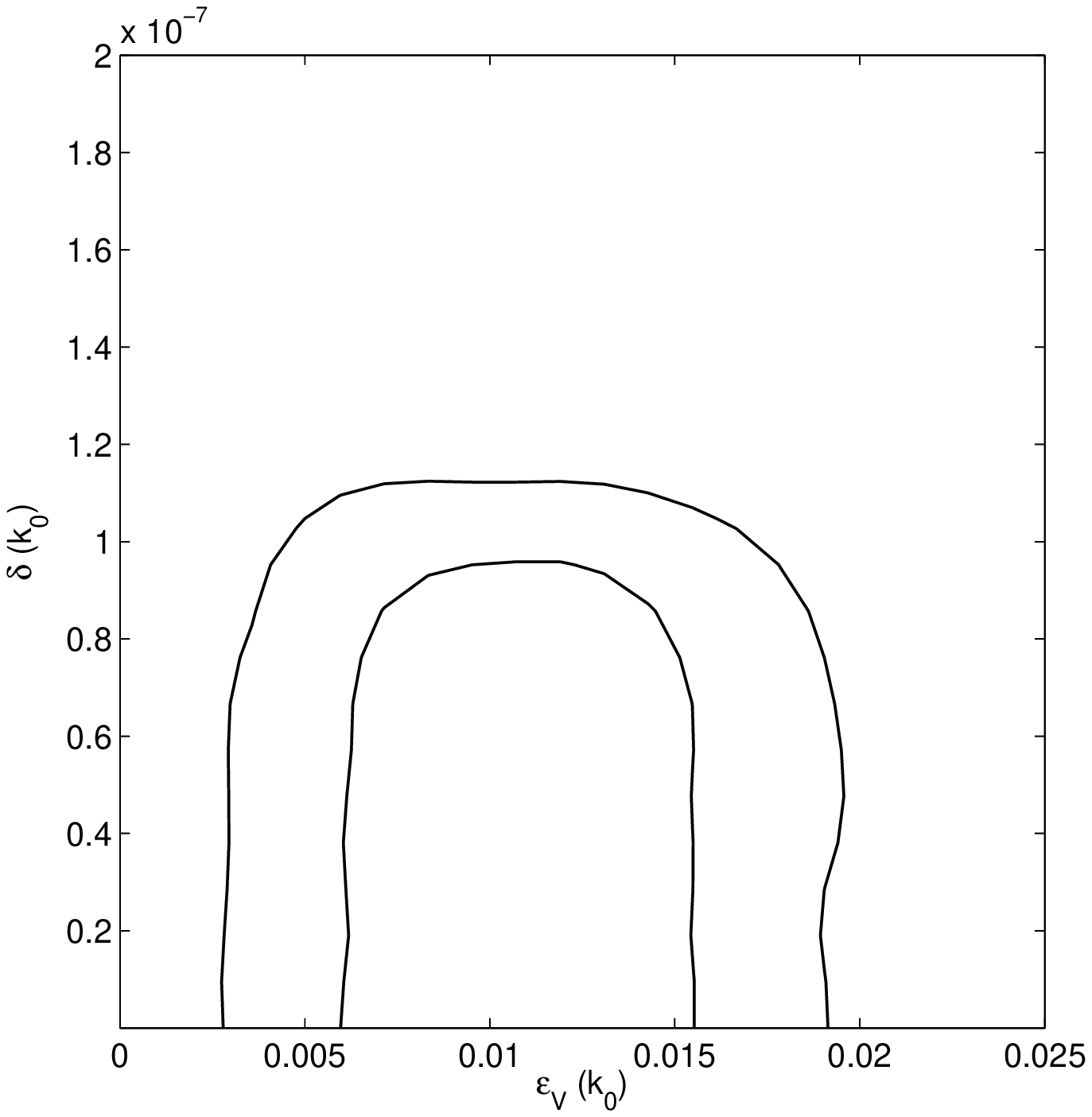}
\caption{2-dimensional marginalized distribution for 
$n=4$ in the two cases: 
(i) $\s=1$ and $k_0=0.002$ Mpc$^{-1}$ (left panel) and 
(ii) $\s=2$ and $k_0=0.05$ Mpc$^{-1}$ (right panel).
\label{en4}}}


Let us proceed to the case of the quartic potential $V(\vp)=V_0 \vp^4$.
Numerically, we find that the observational upper bounds on $\delta (k_0)$
for given $\s$ and $k_0$ are similar to those for the quadratic potential. In the top panel of figure \ref{en4} we show 
the 2-dimensional distribution for $\sigma=1$ with the pivot 
wavenumber $k_0=0.002$ Mpc$^{-1}$.
The LQC correction is constrained to be $\delta (k_0)<
3.4 \times 10^{-2}$ (95 \% CL), which is similar to the bound 
$\delta (k_0)< 3.5 \times 10^{-2}$ for $n=2$ (see table  \ref{tab1}).
The bottom panel of figure \ref{en4} corresponds to the posterior
distribution for $\s=2$ with $k_0=0.05$ Mpc$^{-1}$, in which 
case $\delta (k_0)< 1.1 \times 10^{-7}$ (95 \% CL).
Since the LQC correction given in eq.~(\ref{deltak}) does not depend 
on the values of $n$, the above property of $n$-independence 
can be expected. For larger $\s$ and $k_0$ the upper bounds 
on $\delta (k_0)$ tend to be smaller. 

{}From eq.~(\ref{then}), the values of $\epsilon_\V$ related with the 
CMB anisotropies fall in the range $0.015<\epsilon_\V<0.022$.
For $\s=1$ and $k_0=0.002$ Mpc$^{-1}$, this range is outside the 
1$\sigma$ likelihood contour. In particular, for $N<58$, this model 
is excluded at the 95\% confidence level.
For $\s=2$ the observationally allowed region of $\epsilon_\V$ is 
slightly wider than that for $\s=1$. However, as we see in the lower 
panel of figure \ref{en4}, this model is still under an observational pressure.
Numerically we have confirmed that the bounds on 
$\epsilon_\V(k_0)$ are insensitive to the choice of $k_0$.
Hence the quartic potential is in tension with observations
even in the presence of the LQC corrections.

\subsection{Exponential potentials}

\FIGURE{
\includegraphics[width=7.0cm]{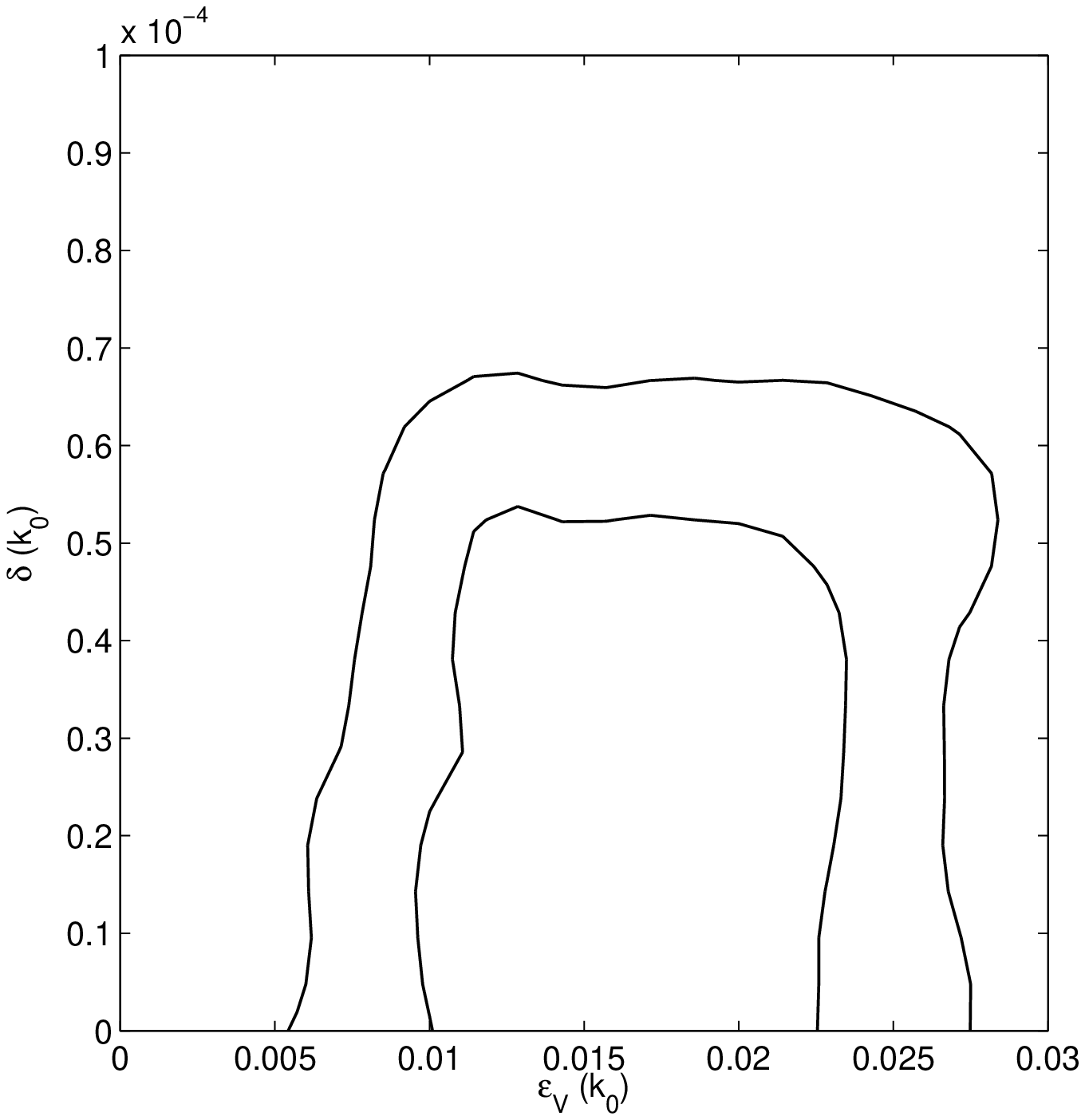}
\includegraphics[width=7.0cm]{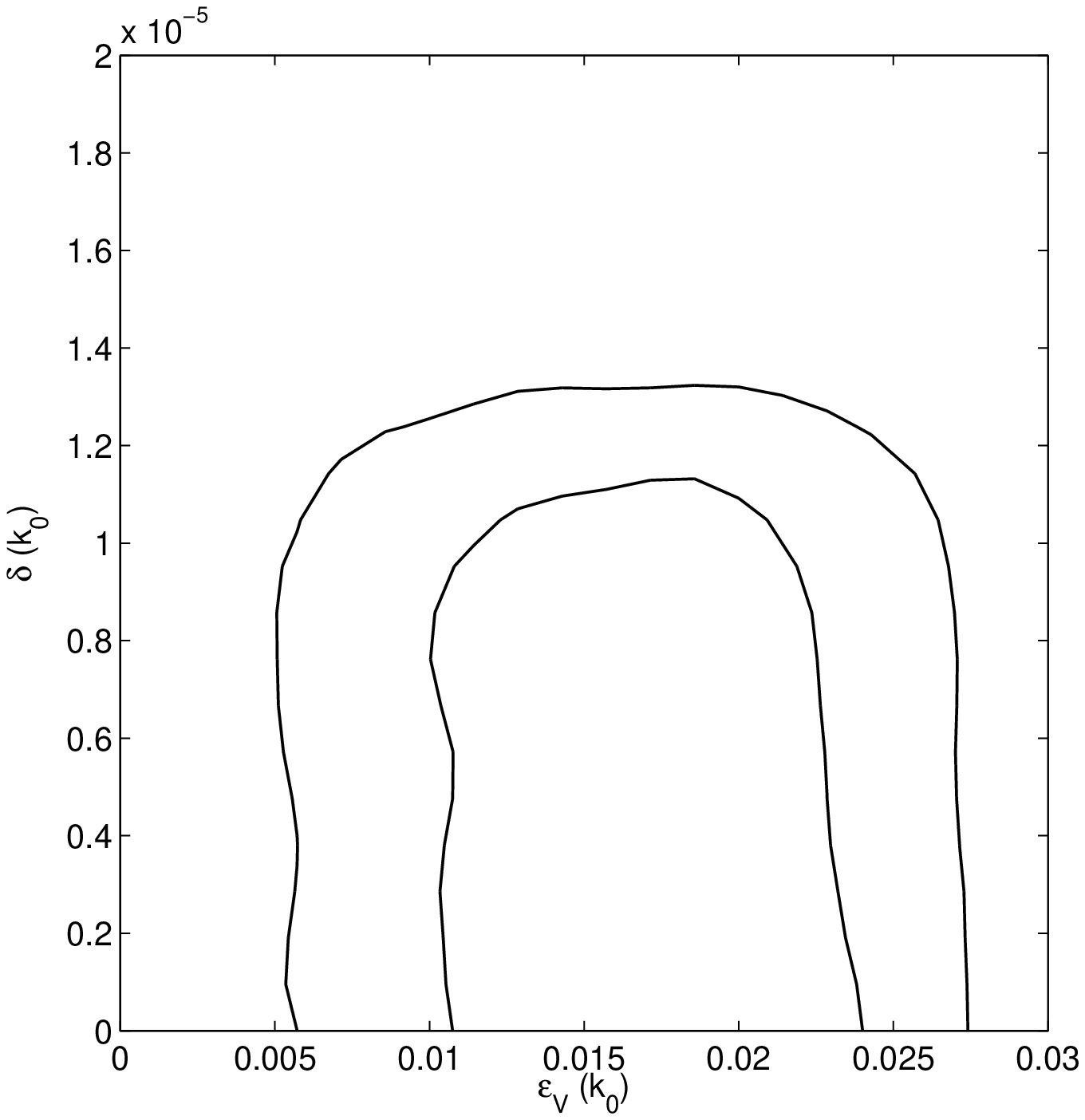}
\caption{2-dimensional marginalized distribution for 
the exponential potential $V(\vp)=V_0 \rme^{-\kappa \lambda \vp}$ 
in the two cases: 
(i) $\s=2$ and $k_0=0.002$ Mpc$^{-1}$ (left panel) and 
(ii) $\s=1.5$ and $k_0=0.05$ Mpc$^{-1}$ (right panel).
\label{expfig}}}


Finally, we study the case of exponential potentials. 
In figure \ref{expfig} we show the 2-dimensional posterior distribution 
for (i) $\s=2$ and $k_0=0.002$ Mpc$^{-1}$ and
(ii) $\s=1.5$ and $k_0=0.05$ Mpc$^{-1}$.
The observational upper limits on the LQC corrections
for the cases (i) and (ii) are $\delta (k_0)<6.8 \times 10^{-5}$
and $\delta (k_0)<1.3 \times 10^{-5}$, respectively, which 
are similar to those for $n=2$ with same values of $\s$.
Hence, for given values of $\s$ and $k_0$, the effect of 
the LQC corrections to the power spectra
is practically independent of the choice of the inflaton potentials.

On the other hand, the observational constraints on the slow-roll
parameter depend on the potential.
In figure \ref{expfig} we find that the observationally allowed values of 
$\epsilon_\V(k_0)$ are in the range $0.005<\epsilon_\V (k_0)<0.27$ 
(95 \% CL) for two different choices of $k_0$.
The maximum value of $\epsilon_\V(k_0)$ is larger than that 
for $n=2$ and $n=4$. Since inflation does not end for 
exponential potentials, one cannot estimate the range of 
the slow-roll parameter relevant to the CMB anisotropies.
Hence one needs to find a mechanism of a graceful exit from inflation
in order to address this issue properly.

\section{Conclusions}

In the presence of the inverse-volume corrections in LQC, we have
provided the explicit forms of the scalar and tensor spectra convenient 
to confront inflationary models with observations.
Even if the LQC corrections are small at the background level, they 
can significantly affect the runnings of spectral indices.
We have consistently included the terms of order higher than the scalar/tensor runnings. 
Inverse-volume corrections generally 
lead to an enhancement of the power spectra at large scales. 

Using the recent observational data of WMAP 7yr combined with LSS, 
HST, SN Ia, and BBN, and analyzing them with techniques routinely used also in standard inflation,
we have placed constraints on the power-law 
potentials $V(\vp)=V_0\vp^n$ ($n=2, 4$) as well as the exponential 
potentials $V(\vp)=V_0 \rme^{-\kappa \lambda \vp}$.
The inflationary observables (the scalar and tensor power spectra 
${\cal P}_{\rm s}$, ${\cal P}_{\rm t}$ and the tensor-to-scalar ratio $r$)
can be written in terms of the slow-roll parameter 
$\epsilon_\V=(V_{,\varphi}/V)^2/(2\kappa^2)$ and
the normalized LQC correction term $\delta$.
We have carried out a likelihood analysis by varying these two parameters 
as well as other cosmological parameters for two pivot wavenumbers $k_0$
($0.002$ Mpc$^{-1}$ and $0.05$ Mpc$^{-1}$).

The observational upper bounds on $\delta (k_0)$ tend to be smaller
for larger values of $k_0$. In table  \ref{tab1} we listed the observational 
upper limits on $\delta (k_0)$ as well as the theoretical priors 
$\delta_{\rm max}$ for the quadratic potential $V(\vp)=V_0\vp^2$
with a number of different values of the quantum gravity parameter $\sigma$
(which is related to $\delta$ as $\delta \propto a^{-\sigma}$).
For larger $\sigma$, we find that $\delta (k_0)$ needs to be suppressed
more strongly to avoid the significant enhancement of the power spectra at large scales.
When $\sigma \lesssim 0.5$ the observational upper limits of $\delta (k_0)$
exceed the theoretical prior $\delta_{\rm max}$, which means that the expansion in terms of 
the inverse-volume corrections can be trustable for $\sigma \gtrsim 0.5$.

As we see in Figs.~\ref{en2}-\ref{expfig} and in table  \ref{tab1}, 
the observational upper bounds on $\delta (k_0)$ for 
given $k_0$ and $\sigma$ are practically independent of 
the choice of the inflaton potentials. This property comes from the fact that the 
LQC correction for the wavenumber $k$ is approximately 
given by $\delta (k)=\delta (k_0) (k_0/k)^{\sigma}$,
which only depends on $k_0$ and $\sigma$.
On the other hand the constraints on the slow-roll parameter
$\epsilon_{\V}$ are different depending on the choice 
of the inflaton potentials.
We have found that the quadratic potential is consistent 
with the current observational data even in the presence of 
the LQC corrections, but the quartic potential is under 
an observational pressure. 
For the exponential potentials the larger values of $\epsilon_{\V}$ 
are favored compared to the power-law potentials.
However, the exponential potentials are not regarded 
as a realistic scenario unless there is a graceful exit from inflation.

The exponential term $e^{-\sigma x}=(k_0/k)^\sigma$ in eq.~\Eq{Psfinal} is responsible for the enhancement of the power spectrum at large scales. This feature is characteristic of the model and is not reproduced by other sources.
For instance, non-commutative geometry or string corrections \cite{Maartens} predict a suppression, rather than an enhancement, of the spectra. Also some old papers on LQC advertized a suppression of power (e.g., the second reference of \cite{inflationworks}), but the quantum corrections were not under full control at the level of perturbation theory; the present results supersede those early discussions. Moreover, the signatures of the LQC spectra cannot be mimicked by any standard scalar potential. The enhancement of power is due to a scale-dependent correction in the spectral amplitudes, while exotic potentials would not affect the perturbed dynamical equations (compare with the LQC Mukhanov equations \cite{InflObs}). In particular, the consistency relation \Eq{tts2} is notably different with respect to the standard relation $r=-8 n_{\rm t}$. Finally, also Wheeler--DeWitt quantum cosmology predicts an enhancement of power at large scales \cite{KiK}, but the effect is qualitatively different from the structure of $\dpl$ and its size is much smaller. The main reason is that it is governed by the energy scale of inflation, contrary to what happens in LQC.

If we compare the observational upper bounds with the theoretical
lower bounds discussed in section \ref{invo}, we can see that estimates
of these parameters are separated by at most a few orders of
magnitude, much less than is usually expected for quantum gravity. By
accounting for fundamental spacetime effects that go beyond the usual
higher-curvature corrections, quantum gravity thus comes much closer
to falsifiability than often granted.  It is of interest to see how
the future high-precision observations such as PLANCK will constrain
the LQC correction as well as the slow-roll parameters. Even in the
case where the quadratic potential were not favored in future
observations, it would be possible that the small-field inflationary
models be consistent with the data.  For these general inflaton
potentials, the effect of inverse-volume corrections on the CMB
anisotropies should be similar to that studied in this paper.

\section*{Acknowledgments}
\label{acknow}
M.B.\ was supported in part by NSF grant 0748336.
S.T.\ was supported by the Grant-in-Aid for Scientific
Research Fund of the JSPS Nos.~30318802.  S.T.\ also thanks financial
support for the Grant-in-Aid for Scientific Research on Innovative
Areas (No.~21111006). 



\end{document}